\documentclass{article}

\usepackage{arxiv}

\usepackage[utf8]{inputenc} 
\usepackage[T1]{fontenc}    
\usepackage{hyperref}       
\usepackage{url}            
\usepackage{booktabs}       
\usepackage{amsfonts}       
\usepackage{nicefrac}       
\usepackage{microtype}      
\usepackage{graphicx}

\usepackage{amssymb}
\usepackage{latexsym}
\usepackage{amsmath}
\usepackage{amsfonts}
\usepackage{physics}
\usepackage{enumitem}
\usepackage{caption}
\usepackage{tabu}
\usepackage{stackengine}
\usepackage{lineno}
\usepackage{graphicx}
\usepackage{floatrow}
\usepackage{subfig}
\usepackage{mathtools}
\usepackage{ulem}  
\usepackage{cancel} 

\usepackage{url}
\usepackage[svgnames,dvipsnames]{xcolor}
\xdefinecolor{dgreen}{named}{DarkOliveGreen}
\definecolor{newcolor}{rgb}{.8,.349,.1}

\usepackage{tikz}
\usetikzlibrary{plotmarks}
\usetikzlibrary{patterns}
\usetikzlibrary{decorations}
\usetikzlibrary{decorations.pathmorphing}
\usetikzlibrary{decorations.pathreplacing}
\usetikzlibrary{decorations.shapes}
\usetikzlibrary{decorations.text}
\usetikzlibrary{decorations.markings}
\usetikzlibrary{decorations.fractals}
\usetikzlibrary{decorations.footprints}
\usetikzlibrary{arrows,calc}
\usetikzlibrary{shapes}

\tikzset{base/.style={draw=gray, align=center, semithick,minimum height=4ex,scale=0.3},
         test1/.style={base, diamond, aspect=3, text width=0.1em, inner sep=1.pt},
        }

\title{Lattice Boltzmann method for computational aeroacoustics on non-uniform meshes: a direct grid coupling approach}

\author{{\hspace{1mm}Thomas Astoul}\\
	CERFACS, 42 Avenue G. Coriolis,\\
	31057 Toulouse Cedex, France\\
	Airbus Operations,316 Route de\\
	Bayonne, 31300 Toulouse, France\\
	\texttt{tastoul@cerfacs.fr}\\
	\And{\hspace{1mm}Gauthier Wissocq} \\
	CERFACS, 42 Avenue G. Coriolis,\\
	31057 Toulouse Cedex, France\\
	\And{\hspace{1mm}Jean-Fran{\c{c}}ois Boussuge} \\
	CERFACS, 42 Avenue G. Coriolis,\\
	31057 Toulouse Cedex, France\\
	\And{\hspace{1mm}Alois Sengissen} \\
	Airbus Operations,316 Route de\\
	Bayonne, 31300 Toulouse, France\\
	\And{\hspace{1mm}Pierre Sagaut} \\
	Aix Marseille Univ, CNRS.\\
	Centrale Marseille, M2P2 UMR 7340,\\
	13451 Marseille, France
}

\renewcommand{\headeright}{}

\begin{document}

\newcommand{\reels}{\mathbb{R}}
\newcommand{\red}{\textcolor{red}}
\newcommand{\orange}{\textcolor{orange}}
\newcommand{\green}{\textcolor{LimeGreen}}
\newcommand{\blue}{\textcolor{blue}}
\newcommand{\purple}{\textcolor{purple}}

\newcommand{\herm}{\mathcal{H}}
\newcommand{\TF}{\begin{tikzpicture} \draw node[circle,draw,thick,fill=blue!40, scale=0.7]{}; \end{tikzpicture}}
\newcommand{\TFN}{\begin{tikzpicture} \draw node[circle,draw,thick,fill=gray!50, scale=0.7]{}; \end{tikzpicture}}
\newcommand{\IE}{\begin{tikzpicture} \draw node[rectangle,draw,thick,fill=gray!50, scale=0.95]{}; \end{tikzpicture}}
\newcommand{\Mid}{\begin{tikzpicture} \node[mark size=5pt,fill=white,thick,scale=0.7] {\pgfuseplotmark{triangle}}; \end{tikzpicture}}

\newcommand{\Intp}{\begin{tikzpicture} \draw node[circle,draw,thick,fill=red!50,scale=0.7]{}; \end{tikzpicture}}

\newcommand{\symbSTD}{\begin{tikzpicture} \draw node[circle,draw=NavyBlue,thick,fill=white, scale=0.55]{}; \end{tikzpicture}}
\newcommand{\symbDCOne}{\begin{tikzpicture} \draw[rotate=0,scale=0.8,thick] (0,0) -- (0.23,0) -- (1/2*0.23, {sqrt(3)/2*0.23}) -- cycle[draw=OliveGreen,fill=white]{};\end{tikzpicture}}
\newcommand{\symbDCTwo}{\begin{tikzpicture} \draw node[rectangle,draw=BrickRed,thick,fill=white, scale=0.7,rotate=45]{}; \end{tikzpicture}}

\newcommand{\lineSTD}{\begin{tikzpicture} \draw [thick,draw=NavyBlue] (-0.25,-0.0) -- (0.25,-0.0); \draw node[circle,draw=NavyBlue,thick,fill=white, scale=0.55]{};  \end{tikzpicture}}
\newcommand{\lineDCOne}{\begin{tikzpicture}  \draw [thick,draw=OliveGreen] (-0.17,0.07) -- (0.33,0.07);\draw[rotate=0,scale=0.8,thick] (0,0) -- (0.23,0) -- (1/2*0.23, {sqrt(3)/2*0.23}) -- cycle[draw=OliveGreen,fill=white]{};\end{tikzpicture}}
\newcommand{\lineDCTwo}{\begin{tikzpicture} \draw [thick,draw=BrickRed] (-0.25,-0.0) -- (0.25,-0.0); \draw node[rectangle,draw=BrickRed,thick,fill=white, scale=0.6,rotate=45]{}; \end{tikzpicture}}

\newcommand{\ArrowDash}{\begin{tikzpicture}  \draw node[circle,draw=white,thick,fill=white, scale=0.6]{}; \draw [-> , >=latex ,line width=0.5mm,dashed] (0,-0.03) -- (0.8,-0.03); \end{tikzpicture}}
\newcommand{\Arrow}{\begin{tikzpicture} \draw node[circle,draw=white,thick,fill=white, scale=0.6]{}; \draw [-> , >=latex ,line width=0.5mm] (0,-0.03) -- (0.8,-0.03); \end{tikzpicture}}
\newcommand{\blackdashPulse}{\begin{tikzpicture}[scale=1] \draw (-0.1,0) node{} ; \draw (0.1,0) node{} ; \draw [thick,draw=black, dotted] (-0.25,-0.03) -- (0.2,-0.03);\end{tikzpicture}}

\newcommand{\redBrickdashed}{\begin{tikzpicture}[scale=1] \draw (-0.12,0) node{} ; \draw (0.12,0) node{} ; \draw [thick,draw=BrickRed,  dashed, line width=1.] (-0.28,0) -- (0.28,0);\end{tikzpicture}}
\newenvironment{Hfigure}{\setcaptiontype{figure}%
  \vskip\textfloatsep\begin{minipage}{\columnwidth}}%
  {\end{minipage}\vskip\textfloatsep\noindent}

\maketitle

\begin{abstract}
The present study proposes a highly accurate lattice Boltzmann direct coupling cell-vertex algorithm, well suited for industrial purposes, making it highly valuable for aeroacoustic applications. It is indeed known that the convection of vortical structures across a grid refinement interface, where cell size is abruptly doubled, is likely to generate spurious noise that may corrupt the solution over the whole computational domain. This issue becomes critical in the case of aeroacoustic simulations, where accurate pressure estimations are of paramount importance. Consequently, any interfering noise that may pollute the acoustic predictions must be reduced.

\noindent The proposed grid refinement algorithm differs from conventionally used ones, in which an overlapping mesh layer is considered. Instead, it provides a direct connection allowing a tighter link between fine and coarse grids, especially with the use of a coherent equilibrium function shared by both grids. Moreover, the direct coupling makes the algorithm more local and prevents the duplication of points, which might be detrimental for massive parallelization. This work follows our first study (Astoul~\textit{et al. 2020}~\cite{Astoul2020}) on the deleterious effect of non-hydrodynamic modes crossing mesh transitions, which can be addressed using an appropriate collision model. The Hybrid Recursive Regularized model is then used for this study. The grid coupling algorithm is assessed and compared to a widely-used cell-vertex algorithm on an acoustic pulse test case, a convected vortex and a turbulent circular cylinder wake flow at high Reynolds number.\\

\end{abstract}


\newpage
\section{Introduction}
\label{sec:intro}
The lattice Boltzmann method (LBM) is an efficient numerical method for simulating complex flows commonly encountered in many fields of physics such as two-phase flows~\cite{Shan1993,Luo1998}, turbulent flows~\cite{Yu2005,Sagaut2010}, or aeroacoustics~\cite{Appelbaum2018}. This numerical method has many advantages making it industrially very attractive. For instance, the locality and the simplicity of the numerical schemes used in the space/time discretization allow a massive parallelization on high-performance computers and offer promising perspectives on GPUs~\cite{Schonherr2011}. Moreover, this method weaves a very close link between its discretization and the topology of the computational mesh. Thus, a Cartesian mesh is mostly used. This type of grid, even though preventing the construction of body-fitted meshes, is very easy to generate. Many industrial LBM solvers directly embed an automatic mesher, which greatly reduces pre-processing turnaround times. For industrial needs, the ability to adapt the mesh resolution when moving away from the areas of flow interest is crucial in order to reduce computation costs. For this purpose, meshes with an octree structure are mostly used~\cite{Ravetta2018}. These meshes are based on refinements of an integer factor between two different resolution domains. 

\noindent The pioneering algorithms designed to couple two Cartesian grids of different sizes within the lattice Boltzmann method framework are respectively those of Filippova and H{\"{a}}nel~\cite{Filippova1998} for cell-vertex type algorithms and Rohde~\cite{Rohde2006} and Chen~\cite{Chen2006} for cell-centered ones. Many algorithms have then been built based on these seminal works. For cell-vertex algorithms, the work of Dupuis and Chopard~\cite{DupuisChopard2003} can be mentioned in which the distribution functions are rescaled before the collision step. Later, Lagrava~\cite{Lagrava2012} or Touil~\cite{Touil2014} added a spatial filtering of the fine distribution functions before transferring them to the coarse mesh, leading to an increase of the numerical stability for turbulent flows. Other families of algorithms were then developed, such as the direct coupling algorithm of Kuwata ~\cite{Kuwata2016a}, that is however restricted to incompressible flows, or algorithms using a finite difference discretization of the discrete velocity Boltzmann equation~\cite{Fakhari2014,Fakhari2015}. 

\noindent These algorithms have been  mostly validated considering purely aerodynamic applications~\cite{Stiebler2011,Touil2014,Dorschner2016} but only few developments dedicated to aeroacoustic applications can be found in the literature. The later are much more demanding since acoustic pressure fluctuations are several orders of magnitude smaller than aerodynamic ones. 
Thus, a very small error in the transmission of an aerodynamic field across a grid interface results in the emission of spurious acoustic waves that may pollute the whole aeroacoustic prediction. A very interesting literature review of aerodynamic and aeroacoustic applications involving non-uniform meshes has been carried out by Gendre \textit{et al.}~\cite{Gendre2017}. The main outcome is that only two studies before his were dealing with aeroacoustics, and only Hasert's phD thesis~\cite{Hasert2014} showed aeroacoustic results at a relatively high Mach number in the presence of a turbulent flow. 
However, strong spurious oscillations are to be deplored in these simulations. Astoul \textit{et al.}~\cite{Astoul2020} have recently highlighted the involvement of the non-hydrodynamic modes inherent to the LBM in these spurious oscillations. In particular, this study has revealed that such modes can generate a significant amount of spurious vorticity, as well as unintended noise.
The effect of these modes was addressed by a careful choice of collision model: the Hybrid Recursive Regulated one (H-RR) \cite{Jacob2018,Feng2019,renard2020improved,guojcp2020}. Once the model in the fluid core is chosen, the quality of the grid coupling algorithm remains essential. For this purpose, Gendre \textit{et al.}~\cite{Gendre2017} proposed an algorithm and validated it considering academic aeroacoustic test cases. It allowed them to greatly reduce spurious noise compared to Lagrava's widely used algorithm~\cite{Lagrava2012}. However, this algorithm is based on ghost cells that make its implementation delicate and may decrease the accuracy of the algorithm for arbitrary shaped transitions. Furthermore, this algorithm was not assessed for turbulent flows representative of typical aeroacoustic applications, such as landing gear noise~\cite{Hou2019}, cavity noise~\cite{Coreixas2015} or air system noise~\cite{Bocquet2019}.

\noindent Under the light of the aforementioned bibliography, current grid coupling algorithms are still not satisfactory for aeroacoustic applications on arbitrary grids. In this framework, the present "aeroacoustically-compliant" algorithm is proposed, which is an extension of Lagrava's work \cite{phDLagrava2012}. In his Ph. D. thesis, Lagrava proposed a direct grid-coupling algorithm validated for a Poiseuille flow. This algorithm was restricted to two dimensional configurations with simple plane transitions, and was far from being usable in an industrial context. The present study aims to improve this algorithm in several ways and validate it in an aeroacoustic framework. It is first extended to three-dimensional meshes and refinement interfaces of arbitrary shape. Furthermore, the algorithm's formulation is simplified and generalized with an efficient numerical resolution method. Finally, the reconstruction method for the distribution functions is improved, making it even more accurate for aeroacoustic applications.
 
\noindent The paper is organized as follows. First of all, key features of the lattice Boltzmann method with the H-RR collision operator are briefly summarized in Sec.~\ref{sec:LBM}. Then, direct coupling (DC) grid refinement algorithms are described in Sec.~\ref{sec:DC_description}. Subsequently, in Sec.~\ref{sec:valid_academic}, numerical validations are performed on academic test cases: an acoustic pulse and a vortex convected across a grid interface. Both test cases are declined with plane and inclined transitions. Afterwards, a validation is performed on an highly turbulent flow around a cylinder with arbitrary grid refinement in Sec.~\ref{sec:valid_turbulent}.

\section{Lattice Boltzmann method with the Hybrid-Recursive Regularized collision operator}
\label{sec:LBM}

The lattice Boltzmann method describes the time and space evolution of the discrete particle distribution functions $f_i(\mathbf{x},t)$, which can be viewed as the probability density of finding fictive particles at location $\mathbf{x}$, at time $t$ and advected at discrete velocities $\boldsymbol{\xi} _i$.
In absence of body-force term, its algorithm can be expressed as 
\begin{equation} \label{eq:LBMequation}
f_i \left(\mathbf{x} + \boldsymbol{\xi} _i , t + 1\right) - f_i\left(\mathbf{x},t\right) = \Omega_i(\mathbf{x},t),
\end{equation}

\noindent where $\Omega_i(\mathbf{x},t)$ is the collision operator. Following the conclusions of a previous study on grid refinement algorithms~\cite{Astoul2020}, the Hybrid-Recursive Regularized (H-RR) \cite{Jacob2018} collision operator will be adopted hereafter. 
This collision model has been chosen for its high stability properties \cite{Feng2019} and since it allows an efficient damping of the  non-hydrodynamic modes \cite{Wissocq2019} which are harmful in the case of non-uniform simulations~\cite{Astoul2020}. In the following, cubic Mach number corrective terms $\boldsymbol{\psi}_i$ will be added to overcome the low symmetry issues of the isothermal lattice.\\

\noindent The H-RR collision model belongs to the regularized collision models family, where distribution functions are regularized before the collision step as proposed by Latt and Chopard \cite{Latt2006}. Regularized distributions functions can be expressed as 

\begin{equation} \label{eq:reguProcedure}
	f_{i}^{reg} \equiv f_{i}^{(0)} + f_{i}^{(1)} + \frac{\boldsymbol{\psi}_i}{2},
\end{equation}

\noindent where $f_i^{(0)}$ is the equilibrium distribution function and $f_i^{(1)}$ the regularized off-equilibrium one. Subsequently, the LBM scheme can be modified to apply the regularized procedure on every distribution function $f_{i}$ during the collision step. By adding cubic Mach correction terms to enhance stability, it becomes

\begin{equation} \label{eq:LBMscheme_reg}
f_{i} \left(\mathbf{x} + \boldsymbol{\xi} _i , t + 1\right)= f_{i}^{(0)} + \left(1 - \frac{1}{\tau}\right) f_{i}^{(1), reg} + \frac{\boldsymbol{\psi}_i}{2},
\end{equation}

\noindent where $\tau$ is the discrete relaxation time of the collision model. Details on the corrective term $\boldsymbol{\psi}_i$ are provided in Appendix A. The equilibrium distribution function is usually approximated using an expansion in Hermite polynomials $\herm_i^{(n)}$ up to an order $N$~\cite{Shan2006,Philippi2006a}

\begin{equation} \label{eq:fEqH}
  f_i^{(0)} = \omega_i \sum_{n=0}^{N}\frac{1}{c_s^{2n}n!} \textit{\textbf{a}}_{0}^{(n)}:\herm^{(n)}_i,
\end{equation}

\noindent where the Gaussian weights $\omega_i$ and the lattice constant $c_s$ are characteristic of the lattice of velocities $\boldsymbol{\xi}_i$. Equilibrium coefficients $\textit{\textbf{a}}^{(n)}_{0}$ are obtained by a projection of the Maxwell-Boltzmann distribution function onto the Hermite polynomials $\herm_i^{(n)}$ \cite{Shan2006} defined as

\begin{equation}
\mathbf{\herm}_i ^{(n)} = \frac{\left(-c_s^2\right)^n}{\omega\left(\boldsymbol{\xi}_i\right)} \nabla _{\boldsymbol{\xi}} ^n w\left(\boldsymbol{\xi}_i\right), \qquad \text{with} \qquad \omega\left(\boldsymbol{\xi}\right) = \frac{1}{\left(2\pi c_s^2\right)^{D/2}} \exp \left( -\frac{||\overline{\xi}|| ^2}{2c_s^2} \right),
\end{equation}

\noindent where $ \nabla _{\boldsymbol{\xi}} ^n$ denotes the $n$th-order gradient tensor obtained by $n$ successive derivations with respect to vector $\boldsymbol{\xi}$ and $D$ is the number of spatial dimensions. \\

\noindent Equilibrium coefficients are expressed as

\begin{align} 
&a_{0}^{(0)} =\rho,\\
&a_{0,\alpha}^{(1)} = \rho u_\alpha,\\
&a_{0,\alpha\beta}^{(2)} = \rho u_\alpha u_\beta + \rho c_s^2 \delta_{\alpha\beta},\\
&a_{0,\alpha\beta\gamma}^{(3)} = \rho u_\alpha u_\beta u_\gamma \delta_{\alpha\beta}.
\label{eq:a0coeffs}
\end{align}

\noindent The D3Q19 lattice is used in this study \cite{Qian1992a}. For this lattice, $c_s=1/\sqrt{3}$, the discrete velocities $\boldsymbol{\xi} _i$ are given by

\begin{equation}
\label{eq:xi_i}
	\boldsymbol{\xi}_i =
	\begin{cases}
	\quad	(0,0,0)   &  i=0,\\
	\quad 	(\pm1,0,0),(0,\pm1,0),(0,0,\pm1)   &  i=1-6,\\
	\quad 	(\pm1,\pm1,0),(\pm1,0,\pm1),(0,\pm1,\pm1)  &  i = 7-18,\\
    \end{cases}
 \end{equation}
 
\noindent and the associated Gaussian weights are

\begin{equation}
\label{eq:w_i}
   	\omega_i = 
    \begin{cases}
    	1/3,  &  i=0,\\
    	1/18, &  i=1-6,\\
    	1/36, &  i=7-18.
    \end{cases}
\end{equation}

\noindent Even if the quadrature order of this lattice theoretically restricts the equilibrium expansion up to the second order ($N=2$), some third-order coefficients can be included to enhance the stability of the numerical scheme~\cite{Wissocq2019,Jacob2018}. This leads to 

\begin{equation} \label{eq:fEq_herm}
\begin{split}
	 f_i^{(0)} = \omega_i \rho &\left[  1 + \frac{\boldsymbol{\xi}_i \cdot \boldsymbol{u}}{c_s ^2} + \frac{1}{2c_s^4} \herm^{(2)}_{i}: \boldsymbol{a}^{(2)}_{0} \right.\\ 
&	 \left.+ \frac{1}{2c_s^6} \left( \herm^{(3)}_{i, xxy}a^{(3)}_{0,xxy} + \herm^{(3)}_{i, xxz}a^{(3)}_{0,xxz} + \herm^{(3)}_{i, xyy}a^{(3)}_{0,xyy} + \herm^{(3)}_{i, xzz}a^{(3)}_{0,xzz} + \herm^{(3)}_{i, yyz}a^{(3)}_{0,yyz} + \herm^{(3)}_{i, yzz}a^{(3)}_{0,yzz}  \right) \right].	
\end{split}
\end{equation}

\noindent Macroscopic quantities of interest, namely the density $\rho$ and the velocity $\mathbf{u}$ in the present athermal case, are defined as the following moments of the distribution function

\begin{equation} \label{eq:moment_rho}
	\rho = \sum_{i} f_i,
\end{equation}

\begin{equation} \label{eq:moment_rhoU}
	\rho \boldsymbol{u} = \sum_{i} \boldsymbol{\xi}_i \, f_i.
\end{equation}

\noindent The regularized off-equilibrium function $f_{i}^{(1)}$ \cite{Malaspinas2015} is also truncated at the third-order. It reads

\begin{equation} \label{eq:fNEq_herm}
\begin{split}
	 f_i^{(1),reg} = \omega_i \rho &\left[  1 + \frac{\boldsymbol{\xi}_i \cdot \boldsymbol{u}}{c_s ^2} + \frac{1}{2c_s^4} \herm^{(2)}_{i}: \boldsymbol{a}^{(2)}_{1} \right.\\ 
&	 \left.+ \frac{1}{2c_s^6} \left( \herm^{(3)}_{i, xxy}a^{(3)}_{1,xxy} + \herm^{(3)}_{i, xxz}a^{(3)}_{1,xxz} + \herm^{(3)}_{i, xyy}a^{(3)}_{1,xyy} + \herm^{(3)}_{i, xzz}a^{(3)}_{1,xzz} + \herm^{(3)}_{i, yyz}a^{(3)}_{1,yyz} + \herm^{(3)}_{i, yzz}a^{(3)}_{1,yzz}  \right) \right],	
\end{split}
\end{equation}

\noindent where $\boldsymbol{a}_1^{(n)}$ are the off-equilibrium expansion coefficients.

\noindent In the H-RR collision model \cite{Jacob2018}, second-order coefficients $\boldsymbol{a}_1^{(2)}$ are obtained thanks to an hybrid computation involving both \textit{projected regularized} (PR) coefficients and a \textit{finite difference} (FD) estimation, yielding

\begin{equation} \label{eq:a12_HRR}
	\boldsymbol{a}^{(2)}_{1} = \sigma \boldsymbol{a}^{(2),\mathrm{PR}}_{1} + (1-\sigma)\boldsymbol{a}^{(2),\mathrm{FD}}_{1}  \qquad \qquad \mathrm{with} \quad [0 \leq \sigma\leq 1].
\end{equation}

\noindent The first term $\boldsymbol{a}^{(2),\mathrm{PR}}_{1}$ is obtained by a projection of the off-equilibrium populations onto second-order Hermite polynomials. The term $\boldsymbol{\psi}_i$ is also involved in the projection, which reads

\begin{equation} \label{eq:a1Proj}
	\textit{\textbf{a}}^{(2),\mathrm{PR}}_{1} = \sum \mathbf{\herm}_i ^{(2)} \left( f_i-f_i^{(0)} + \frac{\boldsymbol{\psi}_i}{2} \right).
\end{equation}
\noindent The second part $\boldsymbol{a}^{(2),\mathrm{FD}}_{1}$ is computed using second-order centered finite differences. This is done through the systematic link, established thanks to a Chapman-Enskog expansion~\cite{Chapman1990}, between off-equilibrium populations $f_i^{(1)}$ and the deviatoric tensor $S_{\alpha \beta} = 1/2\left(\nabla \boldsymbol{u} + \left(\nabla \boldsymbol{u} \right)^T\right)$:

\begin{equation} \label{eq:a1FD}
	\sum_{i}\xi_{i,\alpha} \xi_{i,\beta}f_i^{(1)} \simeq -2 \tau \rho c_s^2 S_{\alpha \beta}\simeq\boldsymbol{a}^{(2),\mathrm{FD}}_{1}.
\end{equation}
\noindent A second-order centered finite difference scheme is used to estimate the components of $\boldsymbol{a}^{(2),\mathrm{FD}}_{1}$ as
\begin{align}
    {a}^{(2),\mathrm{FD}}_{1, \alpha \beta} = - \tau \rho c_s^2 \left( \frac{u_\alpha (\boldsymbol{x} + \boldsymbol{e_\beta}) - u_\alpha (\boldsymbol{x} - \boldsymbol{e_\beta})}{2} + \frac{u_\beta (\boldsymbol{x} + \boldsymbol{e_\alpha}) - u_\beta (\boldsymbol{x} - \boldsymbol{e_\alpha})}{2} \right),
    \label{eq:a1FD_scheme}
\end{align}
where $\boldsymbol{e_\alpha}, \boldsymbol{e_\beta} \in \{ \boldsymbol{e_x}, \boldsymbol{e_y}, \boldsymbol{e_z} \}$ are unitary vectors of the Cartesian coordinate system.
Third-order off-equilibrium coefficients $\boldsymbol{a}_1^{(3)}$ are then computed recursively using Malaspinas'recursive formula~\cite{Malaspinas2015}. In the particular case of $\boldsymbol{a}_1^{(3)}$ coefficients in Eq.~(\ref{eq:fNEq_herm}), it reads

\begin{equation} \label{eq:a13RR}
	\textit{a}^{(3)}_{1,\alpha\alpha\beta} = 2u_\alpha a_{1,\alpha\beta}^{(2)}+u_\beta a_{1,\alpha\alpha}^{(2)}, \qquad \textit{a}^{(3)}_{1,\alpha\beta\beta} = 2u_\beta a_{1,\alpha\beta}^{(2)}+u_\alpha a_{1,\beta\beta}^{(2)}.
\end{equation}

\noindent Finally, the Chapman-Enskog expansion~\cite{Chapman1990} allows linking the relaxation time $\tau$ and the dimensionless kinetic viscosity $\nu$ as
\begin{equation} \label{eq:linkTauNu}
	\nu = c_s^2 \left(\tau - \frac{1}{2} \right).\\
\end{equation}

\noindent In this section, the LBM with a H-RR collision model has been introduced. This model is adopted here for its ability to properly filter out non-hydrodynamic contributions, which can be very harmful at mesh transitions \cite{Astoul2020}. In the next section, the direct grid coupling algorithm is introduced.

\section{Description of the direct coupling algorithm}
\label{sec:DC_description}

\subsection{Rescaling of physical quantities}
\label{subsec:rescaling}
Before introducing the refinement algorithm, specific notions to the LBM for non-uniform grids have to be reminded. Due to the adopted form of dimensionless LBM equations introduced in Sec.~\ref{sec:LBM}, some specific features have to be taken into account.

\noindent Thereafter, any quantity related to the fine mesh or coarse one will be referred to with a $f$ or $c$ superscript respectively. Note that, with the latter conventions, the fine mesh size and time step will be used as non-dimensionalizing parameters. Thus, coarse and fine meshes have respectively a mesh size $\Delta x^c=2\Delta x^f=2$. In the convention of an acoustic scaling \cite{Kruger2015}, the time step $\Delta t$ is imposed by the mesh size, leading to $\Delta t^c=2\Delta t^f=2$. Thus, fine cells are updated twice as often as coarse cells.\\

\noindent Viscosity is the first quantity impacted by this resolution change. Indeed, to ensure a continuity of the Reynolds number~\cite{Filippova1998}, it must be rescaled in the following way:

\begin{equation} \label{eq:rescNu}
	\nu ^f = \frac{\Delta x^c}{\Delta x^f} \nu ^c = 2 \nu ^c,
\end{equation}

\noindent leading to the following relation between the relaxation times

\begin{equation} \label{eq:rescTau}
	\tau ^f = 2 \tau ^c - \frac{1}{2}.
\end{equation}

\noindent Moreover, the equilibrium distribution function is not affected by a resolution change since it depends only on macroscopic variables that are continuous. However, the off-equilibrium part of the distribution function is not continuous since it depends on velocity gradients and the relaxation time $\tau$, as shown by Eq.~(\ref{eq:a1FD}), and therefore must be rescaled. A combination of Eq.~(\ref{eq:a1FD}) and Eq.~(\ref{eq:rescTau}) gives the following rescaling relation

\begin{equation} \label{eq:fneqResc}
	f_i^{(1),f} = \mathrm{R}*f_i^{(1),c},
\end{equation}

\noindent where $\mathrm{R}=0.5\ \tau^f/\tau^c$ is the rescaling factor.\\

\noindent Finally, since they depend on the local mesh size, the last quantities that need to be rescaled for the H-RR model with cubic Mach correction terms introduced in Sec.~\ref{sec:LBM}, are the strain rate tensor $\boldsymbol{S}$

\begin{equation} \label{eq:PijResc}
	S_{\alpha \beta}^{f} =  \frac{S_{\alpha \beta}^{c}}{2},
\end{equation}

\noindent as well as the corrective terms $\boldsymbol{\psi}_i$

\begin{equation} \label{eq:PSIResc}
	\boldsymbol{\psi}_i^{f} =  \frac{\boldsymbol{\psi}_i^{c}}{2}.
\end{equation}

\subsection{Description of the algorithm}
\label{subsec:DC_description}

The proposed grid coupling algorithm is introduced in this sub-section. In the standard collide \& stream algorithm, some populations are missing at the grid interface due to the non-existence of neighbouring nodes with a similar resolution. To reconstruct these populations, most of the existing algorithms use an overlapping area. This strategy is adopted both for the classical cell-centered~\cite{Rohde2006,Chen2006} or cell-vertex~\cite{DupuisChopard2003,Filippova1998} algorithms.\\

\begin{figure}[H]
		\begin{tikzpicture}[scale=1.]
			\draw (1,0) grid[step=2.] (4,4);
			\draw (0,0) grid[step=1.] (2,4);
			\tikzstyle{TC}=[circle,draw,thick,fill=gray!22]
			\tikzstyle{TC0}=[circle,draw,thick,fill=blue!45]
			\tikzstyle{IE}=[rectangle,draw,thick,fill=gray!50,scale=1.35]
			\tikzstyle{IF}=[draw,rectangle,thick,fill=white]
			\tikzstyle{Mid}=[mark size=5pt,fill=white,thick]

			\draw (2,0) node[TC]{};
			\draw (2,4) node[TC]{};
			\draw (2,1) node[IE]{};
			\draw (2,3) node[IE]{};
			\node[Mid] at (1,0) {\pgfuseplotmark{triangle}};
			\node[Mid] at (1,1) {\pgfuseplotmark{triangle}};
			\node[Mid] at (1,2) {\pgfuseplotmark{triangle}};
			\node[Mid] at (1,3) {\pgfuseplotmark{triangle}};
			\node[Mid] at (1,4) {\pgfuseplotmark{triangle}};
			\node[Mid] at (4,0) {\pgfuseplotmark{triangle}};
			\node[Mid] at (4,2) {\pgfuseplotmark{triangle}};
			\node[Mid] at (4,4) {\pgfuseplotmark{triangle}};


			\draw [-> , >=latex ,line width=0.5mm] (2,2) -- (3,3);
			\draw [-> , >=latex ,line width=0.5mm] (2,2) -- (3,2);
			\draw [-> , >=latex ,line width=0.5mm] (2,2) -- (3,1);
			\draw [-> , >=latex ,line width=0.5mm] (2,2) -- (2,1);
			\draw [-> , >=latex ,line width=0.5mm] (2,2) -- (2,3);
			\draw [-> , >=latex ,line width=0.5mm,dashed] (2,2) -- (1,3);
			\draw [-> , >=latex ,line width=0.5mm,dashed] (2,2) -- (1,2);
			\draw [-> , >=latex ,line width=0.5mm,dashed] (2,2) -- (1,1);
			\draw (2,2) node[TC0]{};

			\draw (5,0) grid[step=1.] (9,4);
			\draw [fill=white] (7,0) rectangle (9,2);
			\draw [fill=white] (7,2) rectangle (9,4);
			\draw (7,0) node[TC]{};
			\draw (7,4) node[TC]{};
			\node[Mid] at (9,0) {\pgfuseplotmark{triangle}};
			\node[Mid] at (9,2) {\pgfuseplotmark{triangle}};
			\node[Mid] at (9,4) {\pgfuseplotmark{triangle}};
			\node[Mid] at (6,0) {\pgfuseplotmark{triangle}};
			\node[Mid] at (6,1) {\pgfuseplotmark{triangle}};
			\node[Mid] at (6,2) {\pgfuseplotmark{triangle}};
			\node[Mid] at (6,3) {\pgfuseplotmark{triangle}};
			\node[Mid] at (6,4) {\pgfuseplotmark{triangle}};			
			
			\draw [-> , >=latex ,line width=0.7mm] (7,2) -- (5,4);		
			\draw [-> , >=latex ,line width=0.7mm] (7,2) -- (5,2);
			\draw [-> , >=latex ,line width=0.7mm] (7,2) -- (5,0);
			\draw [-> , >=latex ,line width=0.7mm] (7,2) -- (7,0);
			\draw [-> , >=latex ,line width=0.7mm] (7,2) -- (7,4);
			\draw [-> , >=latex ,line width=0.7mm,dashed] (7,2) -- (9,0);
			\draw [-> , >=latex ,line width=0.7mm,dashed] (7,2) -- (9,2);
			\draw [-> , >=latex ,line width=0.7mm,dashed] (7,2) -- (9,4);		
			\draw (7,2) node[TC0]{};
			
 			\draw (10,3.2) node[IE]{};
 			\node[text width=6.5cm] at(13.7,3.2){Hanging fine nodes without co-located coarse nodes (spatial interpolation)};
 			\draw (10,2.2) node[TC0]{};
 			\draw (10,1.7) node[TC]{};
 			\draw (10.8,1.75) node[TC0]{};
 			\draw (12.95,1.75) node[TC]{};
 			\node[text width=6.5cm] at(13.7,1.95){Co-located coarse and fine nodes \qquad \qquad  \qquad (\quad  \ : Current, \quad \ : Neighbors)};
			\node[Mid] at (10,0.7) {\pgfuseplotmark{triangle}};
 			\node[text width=6.5cm] at(13.7,0.7){Middle nodes};
 			
 			 \node[text width=6.5cm] at(5.05,1.67){0};
 			\node[text width=6.5cm] at(5.33,2.6){3};
 			\node[text width=6.5cm] at(5.5,0.75){7};
 			\node[text width=6.5cm] at(6.1,1.75){1};
 			\node[text width=6.5cm] at(6.1,2.6){2};
 			\node[text width=6.5cm] at(6.1,0.75){8};
 			\node[text width=6.5cm] at(4.3,0.75){6};
  			\node[text width=6.5cm] at(4.3,1.75){5};
 			\node[text width=6.5cm] at(4.3,2.6){4};
 			\node[text width=.5cm] at(7.05,1.67){0};
 			\node[text width=.5cm] at(7.5,3.8){3};
 			\node[text width=.5cm] at(7.5,0.2){7};
 			\node[text width=.5cm] at(9.,1.65){1};
 			\node[text width=.5cm] at(8.7,3.8){2};
 			\node[text width=.5cm] at(8.7,0.2){8};
 			\node[text width=.5cm] at(5.7,0.2){6};
  			\node[text width=.5cm] at(5.3,1.65){5};
 			\node[text width=.5cm] at(5.7,3.8){4};
 			
		\end{tikzpicture}
	\caption{\label{fig:scheme_DC} Two dimensional representation of a plane refinement interface. (\protect\ArrowDash): Unknown distribution functions after a streaming step, (\protect\Arrow): known distribution functions. Left: fine domain, right: coarse domain.}
\end{figure}
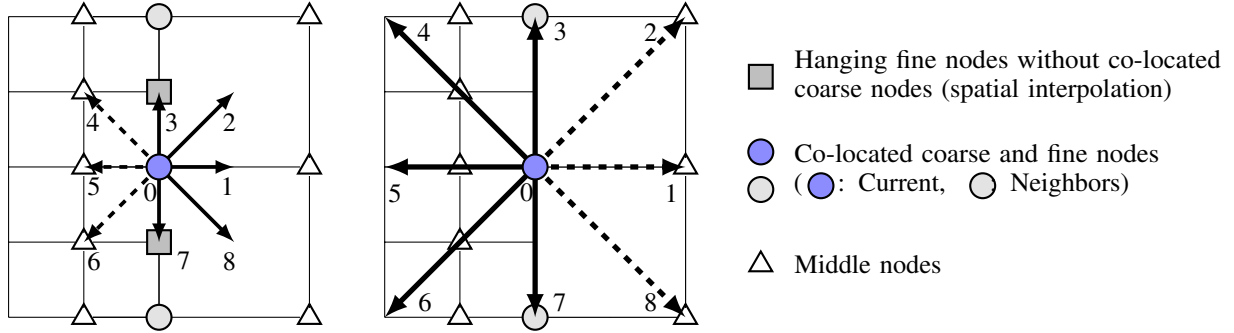

\noindent The present algorithm does not require any overlapping area to achieve the grid coupling. It only needs some particular treatments performed on two specific nodes (\TF, \IE). Both of these nodes are displayed on Fig.~\ref{fig:scheme_DC}. The first ones are co-located fine and coarse nodes. They are used to reconstruct the missing populations on both grids. The second ones are hanging fine nodes that do not have any counterpart in the coarse domain.

\noindent At the grid interface, several distribution functions are missing and cannot be streamed on (\TF) nodes (\textit{cf.} Fig.~\ref{fig:scheme_DC}). They have to be reconstructed after the streaming step. The present grid coupling algorithm aims at reconstructing the missing populations based on the following hypothesis ensuring mass and momentum conservation at (\TF) nodes

\begin{equation} \label{eq:cond_fneq}
   \sum_i \boldsymbol{\Phi}_i {f}_i^{(1), c} = \sum_i \boldsymbol{\Phi}_i {f}_i^{(1), f} = \boldsymbol{0},
\end{equation}

\noindent with $\boldsymbol{\Phi}_i=(1,\xi_{x,i},\xi_{y,i},\xi_{z,i})^\mathrm{T}$.\\

\noindent In practice, this equality cannot be satisfied straightforwardly since distributions with the same resolution are missing in each grid. Eq.~(\ref{eq:fneqResc}) can be used to relate the off-equilibrium distributions of both grids. It is therefore possible to ensure Eq.~(\ref{eq:cond_fneq}) by mixing the off-equilibrium distribution functions belonging to one mesh with the other. Within the off-equilibrium part, $f_i$ can be known since it corresponds to a post-collision function of an existing neighboring node. However, no $f_i^{(0)}$ is known since their computation involves unknown macroscopic quantities. It is with the objective of determining a consistent equilibrium function $f_i^{(0)}$ at (\TF) nodes that the system proposed by the Eq.~(\ref{eq:cond_fneq}) must be solved.\\

\noindent Since fine cells are updated twice as often as coarse ones, the system will be solved in the fine mesh, \textit{i.e.} only the right hand side part of Eq.~(\ref{eq:cond_fneq}) will be considered: $\sum_i \boldsymbol{\Phi}_i f_i^{(1), f}=\boldsymbol{0}$ . Solving this system allows finding the macroscopic variables that are required to compute a local equilibrium on the (\TF) nodes. This equilibrium, in addition to the known distribution functions, will make it possible to reconstruct both missing fine and coarse distribution functions. \\

\noindent For a sake of generality, the set of indexes of fine and coarse missing populations will be respectively referred to as $\mathcal{M}^f$ and $\mathcal{M}^c$. The set of population indexes which are both known on the fine and coarse mesh after the streaming step will be referred to as $\mathcal{P}$, and the set of populations indexes that are only known on the fine mesh (resp. the coarse mesh) will be referred to as $\mathcal{Q}^f$ (resp. $\mathcal{Q}^c$).  For instance, in the particular case of Fig.~\ref{fig:scheme_DC}, one has $\mathcal{M}^f=\mathcal{Q}^c=\{4, 5, 6\}$, $\mathcal{M}^c=\mathcal{Q}^f=\{1, 2, 8\}$ and $\mathcal{P}=\{0, 3, 7\}$. With these notations,  Eq.~(\ref{eq:cond_fneq}) can be re-written for the fine mesh, after replacing the missing fine populations by the rescaled coarse ones thanks to Eq.~(\ref{eq:fneqResc}), as
\begin{align} \label{eq:cond_fneq_sets}
    \sum_{i \in \mathcal{Q}^f} \boldsymbol{\Phi}_i f_i^{(1), f} + \sum_{i \in \mathcal{M}^f} R \cdot \boldsymbol{\Phi}_i f_i^{(1), c} + \sum_{i \in \mathcal{P}} \boldsymbol{\Phi}_i f_i^{(1), f} = \boldsymbol{0}. 
\end{align}
Furthermore, it can be noticed that the last term of this equation, involving $\mathcal{P}$, can either be computed thanks to the fine populations $f_i^{(1), f}$ or their coarse counterpart $f_i^{(1), c}$ after a rescaling by R. For this reason, and in order to generalize the resolution procedure, two parameters $\Gamma_i$ and $\gamma_i$ are introduced below, depending on the nature (\textit{i.e.} fine or coarse) of the distribution functions that are taken into account in the system resolution. Eq.~(\ref{eq:cond_fneq_sets}) can then be re-written as
\begin{equation} \label{eq:cond_fneq_Gamma}
     \sum_{i} \Gamma_i \cdot \boldsymbol{\Phi}_i {f}_i^{(1), \gamma_i} = \boldsymbol{0}, 
\end{equation}
where
\begin{equation} \label{eq:Gamma_def}
   \begin{cases}
    	\text{Fine distribution is used:} & \Gamma_i = 1, \qquad \gamma_i = f,\\
    	\text{Coarse distribution is used:} & \Gamma_i = \mathrm{R}, \qquad \gamma_i=c,
   \end{cases}
\end{equation}

\noindent and where $\mathrm{R}$ is the rescaling factor to convert a coarse to a fine non-equilibrium distribution. The choice of the couple $(\Gamma_i, \gamma_i)$ is not unique and will be further discussed in Sec.~\ref{subsec:Gamma_choice}. 



\noindent In Eq.~(\ref{eq:cond_fneq_Gamma}), off-equilibrium distribution functions $f_i^{(1), \gamma_i} = f_i^{\gamma_i} - f_i^{(0)}(\boldsymbol{\mathrm{X}})$ depend on a vector of macroscopic variables $\boldsymbol{\mathrm{X}}=(\rho,u_x,u_y,u_z)$. This system of equations can be rewritten in the following manner:

\begin{equation} \label{eq:system_fi}
     \boldsymbol{\mathrm{F}}(\boldsymbol{\mathrm{X}}) = \sum_{i} \Gamma_i \cdot \boldsymbol{\Phi}_i \left(f_i^{\gamma_i} - f_i^{(0)}(\boldsymbol{\mathrm{X}}) \right)= \boldsymbol{0}. 
\end{equation}


\noindent In this system, all distribution functions $f_i^{\gamma_i}$ are known, while all equilibrium functions $f_i ^{(0)}(\boldsymbol{\mathrm{X}})$ are unknown. For three dimensional cases, this system contains four equations and four unknowns ($\rho,u_x,u_y,u_z$). Because of the quadratic, cubic or even higher-order powers in velocity arising in the adopted form of equilibrium function $f_i^{(0)}$, this system is genuinely non-linear. Furthermore, the non-equilibrium functions are multiplied by the $\Gamma_i$ parameter, which takes as many values (1 or R) as the number of discrete velocities. For these reasons, the system can be very tough to solve for three dimensional cases and arbitrary grid refinement interfaces.\\

\noindent A general methodology of resolution is proposed here using an iterative Newton-Raphson method~\cite{Kerst1946}. This efficient method allows finding the roots of a given set of equations, here $\boldsymbol{\mathrm{F}}(\boldsymbol{\mathrm{X}}) = \boldsymbol{0}$.\\

\noindent Firstly, one can linearly evaluate, through a Jacobian matrix $\boldsymbol{\mathrm{J_F}}(\boldsymbol{\mathrm{X}_0})=\mathrm{d}\boldsymbol{\mathrm{F}}(\boldsymbol{\mathrm{X}_0})/\mathrm{d}\boldsymbol{\mathrm{X}}$, the value of $\boldsymbol{\mathrm{F}}(\boldsymbol{\mathrm{X}_0}+\delta {\boldsymbol{\mathrm{X}}})$ that is the value of $\boldsymbol{\mathrm{F}}(\boldsymbol{\mathrm{X}_0})$ plus a small variation $\delta {\boldsymbol{\mathrm{X}}}$ around a first estimation of the roots $\boldsymbol{\mathrm{X}_0}$:

\begin{equation}
\boldsymbol{\mathrm{F}}(\boldsymbol{\mathrm{X}_0}+\delta {\boldsymbol{\mathrm{X}}}) \simeq
\boldsymbol{\mathrm{F}}(\boldsymbol{\mathrm{X}_0}) + \boldsymbol{\mathrm{J_F}}(\boldsymbol{\mathrm{X}_0}) \cdot \delta {\boldsymbol{\mathrm{X}}}.
\end{equation}

\noindent Subsequently, by assuming $\boldsymbol{\mathrm{F}}(\boldsymbol{\mathrm{X}_0}+\delta {\boldsymbol{\mathrm{X}}})=\boldsymbol{0}$, the roots $\boldsymbol{\mathrm{X}_0}+\delta {\boldsymbol{\mathrm{X}}}$ can be determined through a linear interpolation:

\begin{equation}
\delta {\boldsymbol{\mathrm{X}}} = \boldsymbol{\mathrm{J_F}}^{-1}(\boldsymbol{\mathrm{X}_0}) \left[ \boldsymbol{\mathrm{F}}(\boldsymbol{\mathrm{X}_0}+\delta {\boldsymbol{\mathrm{X}}}) - \boldsymbol{\mathrm{F}}(\boldsymbol{\mathrm{X}_0}) \right] = -\boldsymbol{\mathrm{J_F}^{-1}}(\boldsymbol{\mathrm{X}_0}) \boldsymbol{\mathrm{F}}(\boldsymbol{\mathrm{X}_0}).
\end{equation}

\noindent The roots can be found from any starting point $\boldsymbol{\mathrm{X}_0}$ as

\begin{equation}
\boldsymbol{\mathrm{X}_0} + \delta {\boldsymbol{\mathrm{X}}}  = \boldsymbol{\mathrm{X}_0} -\boldsymbol{\mathrm{J_F}}^{-1}(\boldsymbol{\mathrm{X}_0}) \boldsymbol{\mathrm{F}}(\boldsymbol{\mathrm{X}_0}).
\end{equation}

\noindent Finally, since $\boldsymbol{\mathrm{F}}$ is nonlinear, the macroscopic variables can be obtained iteratively in the following way:

\begin{equation}
\label{eq:updated_macro}
\forall n \geq 0, \qquad \boldsymbol{\mathrm{X}}_{n+1} = \boldsymbol{\mathrm{X}}_n - \boldsymbol{\mathrm{J_F}}^{-1}(\boldsymbol{\mathrm{X}}_n) \cdot \boldsymbol{\mathrm{F}}(\boldsymbol{\mathrm{X}}_n).
\end{equation}

\noindent The iterative method can be considered as converged when $||\delta \boldsymbol{X}||<10^{-12}$.\\

\noindent Since the system to be solved is nonlinear and depends on the local shape of the interface, the resolution is performed by means of the free formal computation software Maxima~\cite{Maxima}. It must be resolved once and for all before it gets implemented. Practically, the formal computation software provides an analytical expression for the Jacobian matrix $\boldsymbol{\mathrm{J_F}}$. Then the inversion of the Jacobian matrix $\boldsymbol{\mathrm{J_F}}^{-1}$ is performed directly in the LBM code with the LAPACK library~\cite{Lapack} as well as the resolution of the Newton's iterative algorithm. In all the cases of Sec.~\ref{sec:valid_academic}-\ref{sec:valid_turbulent}, the algorithm takes less than three iterations to converge. 

\noindent Once the macroscopic variable at (\TF) nodes are updated, it is then possible to reconstruct the fine missing populations~\cite{DupuisChopard2003} using the new equilibrium distribution function $f_i^{(0)}$ determined with the updated macroscopic variables:
\begin{equation} \label{eq:rescff}
	\forall i \in \mathcal{M}^f, \qquad f_i ^f = f_i ^{(0)} + \mathrm{R} \ f_i ^{(1),c}.
\end{equation}

\noindent Non-coincident nodes (\IE) are completed by means of spatial interpolations. Fourth-order one-dimensional interpolation schemes \cite{Lagrava2012} are used to reconstruct their missing populations. Spatial interpolations are a critical subject for the quality of transition algorithms. It has been shown \cite{Lagrava2012} that at least third-order spatial interpolations are required to ensure mass conservation.
In the present work, the choice of one-dimensional interpolations is adopted for a sake of simplicity and to preserve the computational efficiency of the algorithm. A systematic use of three-dimensional interpolations that depend on the shape of the interface is indeed not conceivable. This choice of one-dimensional interpolation has a significant impact on the quality of simulations, which will be quantified in Sec.~\ref{sec:valid_academic}.\\

\noindent Furthermore, as two fine iterations are performed during one coarse time step, a temporal interpolation of $f_i^c$ is needed to reconstruct the missing populations $f_i^f$ on ($\TF^f$) nodes. A third-order polynomial interpolation is used. For a quantity $g$, it reads:

\begin{equation} \label{eq:TempInt}
g \left( \TF , t + \Delta t^f \right) = -\frac{1}{8} g \left( \TF , t-\Delta t^f \right) + \frac{3}{4} g \left( \TF , t\right) + \frac{3}{8} g \left( \TF , t + \Delta t^c \right).
\end{equation}

\noindent It is noteworthy that no spatial filtering is used when transferring distributions from the fine mesh to the coarse one. Indeed, the algorithm being without overlap, it is not possible to use an isotropic filtering as generally done with cell-vertex algorithms \cite{Lagrava2012,Touil2014}. No particular needs for filtering have been observed in the numerical experiments of Sec.~\ref{sec:valid_academic}-\ref{sec:valid_turbulent}.\\

\noindent A specificity related to the H-RR algorithm remains to be clarified: it involves the computation  of macroscopic gradients to estimate the strain rate tensor $\boldsymbol{S}$ and the corrective term $\boldsymbol{\psi_i}$. Since these gradients are required in the computation of $\boldsymbol{a}^{(2)}_{1}$, which behaves like a diffusive term, a standard centered interpolation is generally preferred. Discretizing them with an off-centered scheme may indeed lead to instabilities as well as loss of accuracy. However, on the co-located nodes of the interface, the gradient estimation with a centered finite difference scheme is not straightforward. Details regarding their computation are given in the following section.

\subsection{Details on gradient computation at the interface}
\label{subsec:grad_computation}

\noindent The present collision model involves the estimation of macroscopic gradients using finite difference schemes (Eqs.~(\ref{eq:a1Proj})-({\ref{eq:a1FD}})). To ensure consistency with the numerical scheme used in the whole fluid domain, it is preferable to compute these gradients at the interface with centered second-order schemes instead of a degraded first-order upwind scheme. This is all the more true since diffusive terms are usually calculated with centered schemes for stability and accuracy reasons.

\begin{figure}[H]
		\begin{tikzpicture}[scale=0.8]
			\draw (1,0) grid[step=2.] (4,4);
			\draw (0,0) grid[step=1.] (2,4);
			\tikzstyle{TC0}=[circle,draw,thick,fill=red!40,scale=1]
			\tikzstyle{IE}=[rectangle,draw,thick,fill=gray!50]

			\draw (2,0) node[IE]{};
			\draw (2,4) node[IE]{};
			\draw (0,2) node[IE]{};
			\draw (4,2) node[IE]{};
			\draw (2,2) node[TC0]{};

		\end{tikzpicture}
	\caption{\label{fig:scheme_gradC} Two dimensional representation of a plane refinement interface. Nodes used for the estimation of the centered second order finite difference gradient at the interface: (\protect\Intp): node that requires the gradient computation, (\protect\IE): nodes used to estimate the gradient}
\end{figure}
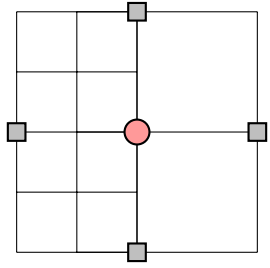

\noindent The standard configuration of a co-located node is displayed on Fig.~\ref{fig:scheme_gradC}. The difficulty here is that there is no fine node placed on the right of the (\Intp) node, which would yet be required to compute the gradient. A coarse stencil is thus necessary.\noindent

\noindent During even iterations, when the two meshes are synchronized with each other, gradients can be easily estimated with a coarse stencil using either the fine or coarse macroscopic variables. Then, they can be transferred to the fine mesh after being rescaled to the fine scale with Eqs.~(\ref{eq:PijResc})-(\ref{eq:PSIResc}).

\noindent However, at odd iterations, macroscopic variables are unknown in the coarse mesh. They have to be temporally interpolated using Eq.~(\ref{eq:TempInt}). These gradients are also estimated with a coarse stencil, they thus need to be rescaled to the fine scale using Eqs.~(\ref{eq:PijResc})-(\ref{eq:PSIResc}).\\

\noindent In the following section, the choice of the $\Gamma_i$ parameter is discussed. This parameter depends on the choice of distributions used in the DC algorithm which can be multiple.

\subsection{Choice of the $\Gamma_i$ parameter and distributions used in the DC algorithm}
\label{subsec:Gamma_choice}

At the grid interface, several possibilities may exist to reconstruct the distribution functions. Populations for which $i \in \mathcal{P}$ are indeed known on both meshes. Other populations for which $i \in \mathcal{Q}^f$ (resp. $\mathcal{Q}^c$) are only known on the fine (resp. coarse) mesh.\\

\noindent This observation leads us to several reconstruction possibilities. All options have been considered in this work and two of them are summarized in Table.~\ref{tab:choice_Gamma_fine}. The first one, referred to as DC1, is a generalization of the reconstruction originally proposed by Lagrava~\cite{phDLagrava2012}. The second one, referred to as DC2, is an improved reconstruction that is proposed in this study.\\

\noindent The distribution functions for the fine mesh reconstruction on  $(\TF)^f$ node and the associated value of $\Gamma_i$ are summarized in Table~\ref{tab:choice_Gamma_fine}.


		


\begin{table}[H]
	\centering
	\begin{tabu}{ |[1.2pt] c|[1.2pt]c|[1.2pt]c|[1.2pt]c|[1.2pt]}
\tabucline[1.2pt]{-}
		Set of indexes  & $\mathcal{P}$ & $\mathcal{Q}^f$ & $\mathcal{M}^f$ \\\tabucline[1.2pt]{-}
		$\mathrm{DC_1}$  & ($1$, f) & ($1$, f) & ($R$, c)\\
\hline

		$\mathrm{DC_2}$  & ($R$, c) & ($1$, f) & ($R$, c) \\
		
\tabucline[1.2pt]{-}

	\end{tabu}
	\caption{\label{tab:choice_Gamma_fine} Couples $(\Gamma_i, \gamma_i$) assigned on $(\protect\TF)$ nodes for the iterative resolution of Eq.~(\ref{eq:system_fi}), according to the set of population indexes $i$.}
\end{table}

\noindent With regard to the reconstruction of the coarse distribution functions $f_i^c$ on $(\TF)^c$ nodes, in the same way as for the fine distribution reconstruction, many possibilities exist. However, a lack of consistency between reconstruction in the fine mesh and in the coarse one may lead to a non-conservation of mass and momentum in the collision step. It is therefore decided here to reconstruct all coarse functions using the previously completed fine ones, as

\begin{equation}
    \label{eq:rescfc}
	\qquad f_i ^c = f_i ^{(0)} + 1/\mathrm{R}\  f_i ^{(1),f}.
\end{equation}

\noindent In summary, the DC1 model reconstructs the fine distribution functions using as many informations coming from the fine mesh as possible. On the contrary, the DC2 model uses as many available coarse functions as possible. The DC2 formulation aims to reduce the reintroduction of interpolation errors occuring on (\IE) nodes in the reconstruction of the (\TF) nodes so as to minimize aliasing effects.\\

\noindent If the interface is not planar as in Fig.~\ref{fig:scheme_DC}, the only difference lies in the indexes associated with the sets $\mathcal{P},\mathcal{Q}^f,\mathcal{M}^f$, and thus the values of associated couple $(\Gamma_i, \gamma_i)$. The system~(\ref{eq:system_fi}) being solved for discrete $(\Gamma_i, \gamma_i)$ values, it is simply enough to substitute these values in the LBM code, by the one corresponding to the given interface. Appendix~B provides some examples of sets $\mathcal{P},\mathcal{Q}^f,\mathcal{M}^f$ for interfaces with corners in two dimensions.\\

\noindent The next section details the different steps of the algorithm.

\subsection{Steps of the algorithm}
\label{subsec:algo_steps}

\noindent The steps of the algorithm are summarized as follows:

\begin{enumerate}[label=\arabic*)]
	\setlength\itemsep{-0.5em}
	\item \textbf{Reference state} $\rightarrow$ Fine grid $t$ ; Coarse grid $t$
	\begin{enumerate}[label=\alph*.]

		\item All the distribution functions are known on both grids.
	\end{enumerate}

	\item \textbf{Asynchronous iteration} $\rightarrow$ Fine grid $t+\Delta t^f$ ; Coarse grid $t+\Delta t^c$ 
	\begin{enumerate}[label=\alph*.]
		\setlength\itemsep{-0.0em}
		\item Propagation step towards fine and coarse middle nodes.
		\item Streaming of fine known populations towards $(\TF)$ nodes.
		\item Streaming of coarse known populations towards $(\TF)$ nodes.
		\item Temporal interpolation of previously streamed coarse populations $f_{i}^c$ on $(\TF^f)$ nodes using Eq.~(\ref{eq:TempInt}).
		\item Reconstruction of missing fine populations with parameter $\Gamma_{i}=\mathrm{R}$ with Eq.~(\ref{eq:rescff}). $f_i^{(0)}$ is deduced from the macroscopic variables obtained thanks to the iterative scheme of Eq.~(\ref{eq:updated_macro}).
		\item Estimation of the fine strain tensor $S^f_{\alpha\beta}$ and the cubic Mach corrective term $\boldsymbol{\psi}^f$ on $(\TF^f)$ nodes using a second-order centered interpolation scheme following the methodology introduced in Sec.~\ref{subsec:grad_computation}.
		\item Spatial interpolation of $f_{i}^f$, $\boldsymbol{\psi}^f$ and $S^f_{\alpha\beta}$ on $(\IE)$ nodes. 
		\item Collision of all fine nodes.
		\end{enumerate}
	
	\item \textbf{Synchronous iteration} $\rightarrow$ Fine grid $t+\Delta t^c$ ; Coarse grid $t+\Delta t^c$
	\begin{enumerate}[label=\alph*.]
		\setlength\itemsep{-0.em}
		\item Propagation step towards the fine middle nodes.
		\item Streaming of fine known populations towards $(\TF^f)$ nodes
		\item Reconstruction of missing fine populations with parameter $\Gamma_{i}=\mathrm{R}$ with Eq.~(\ref{eq:rescff}). The coarse populations used here are those streamed on step 2c. $f_i^{(0)}$ is deduced from the macroscopic variables obtained thanks to the iterative scheme of Eq.~(\ref{eq:updated_macro}).
		\item Estimation of the coarse strain tensor $S^c_{\alpha\beta}$ and the cubic Mach correction term $\boldsymbol{\psi}^c$ on $(\TF^c)$ using second-order centered interpolation scheme. On $(\TF^f)$ nodes, transfer and conversion of $S^c_{\alpha\beta}$ and $\boldsymbol{\psi}^c$ to the fine scale using Eq.~(\ref{eq:PijResc}) and Eq.(~\ref{eq:PSIResc}) respectively.
		\item Spatial interpolation of $f_{i}^f$, $S^f_{\alpha\beta}$ and $\boldsymbol{\psi}^f$ on $(\IE)$ nodes.
		\item Reconstruction of coarse populations using (\ref{eq:rescfc}).
		\item Collision of all nodes with Eq.~(\ref{eq:LBMscheme_reg}).
	\end{enumerate}
	\item \textbf{Repetition of steps 2) to 4) until the end of the simulation}.\\
\end{enumerate}

\noindent This grid refinement algorithm is made completely generic for any orientation of the interface and either two or three dimensional configurations. The only difficulty lies in the preliminary needs to use a formal computing tool to solve the system determining the equilibrium function. Using a direct connection like this one allows reducing the number of duplicated points at grid interface compared to overlapping algorithms, which saves memory and improves code parallelization.\\

\noindent All the useful theoretical details on the H-RR collision model and the grid coupling algorithm being described, the validation of the DC algorithms are presented in the following sections.


\section{Numerical validation and comparison with existing grid refinement algorithm on academic test cases}
\label{sec:valid_academic}

In this section, numerical validations and comparisons with a standard cell-vertex algorithm will be performed. Firstly, a two-dimensional acoustic pulse is considered over a plane and a circular interface. Secondly, a convected vortex through a plane and an inclined grid refinement interface is investigated.

\noindent Results will be confronted with the standard cell-vertex algorithm with overlapping area described in~\cite{Lagrava2012,Touil2014}, which was also presented and assessed in our previous study \cite{Astoul2020}. The fine to coarse filtering used is the one presented in~\cite{Touil2014}. This algorithm will be referred to as STD. It is adopted as a reference since it is widely used in the literature~\cite{Brogi2017,Touil2014,Sengissen2015,Hou2019}.

\noindent All the simulations of the present study are carried out with a kinematic viscosity $\nu=1.49.10^{-5}\ m^2.s^{-1}$, a speed of sound $c_0=347.3\ m.s^{-1}$ and an hybridization parameter $\sigma=0.98$. 

\subsection{Acoustic Pulse}
\label{subsec:acous_pulse}

A pseudo-2D acoustic pulse is considered in this section. This is a purely acoustic test case.
The acoustic pulse is initialized in the fine grid as follows:

\begin{align}
\label{eq:init_Pulse}
		& \quad \rho \left(x,y,z\right) = \rho_0\left( 1 + A \exp  \left(- \frac{(x+y)^2}{2R_c^2}\right)\right), \\
	 	& \quad \mathbf{u}\left(x,y,z\right) = \mathbf{0},
\end{align}
with
\begin{equation}
	\label{eq:init_Pulse2}
	\begin{split}
		\quad \rho_0 = 1\ kg.m^{-3} \ ,
   		\quad A = 10^{-3},
		\quad R_c = 0.1\ m.
\end{split}
\end{equation}

\noindent Distribution functions are initialized as their equilibrium value (\textit{cf.} Eq.~(\ref{eq:fEq_herm})) computed with these macroscopic quantities. The simulated domain is a pseudo 2D periodic box of size [$L$,$L$,$\Delta x^c$] with $L=3m$ and $\Delta x^c=2\Delta x^f=0.02m$. The pulse is initialized at the center of the box.

\subsubsection{Acoustic pulse across a plane refinement interface}
\label{subsubsec:pulse_plane}

In this section, propagation of the acoustic pulse across a plane interface is considered. The grid interfaces are located at $x=-0.75m$ and $x=0.75m$. The computational domain is sketched on Fig.~\ref{fig:Pulse_domain}.

\begin{minipage}[l]{0.45\linewidth}
\begin{figure}[H]
	\begin{center}
		\begin{tikzpicture}[scale=0.78]
			\draw (0.,0) grid[step=0.3] (1.5,6);
			\draw (1.5,0) grid[step=0.15] (4.5,6.001);
			\draw (4.5,0) grid[step=0.3] (6.,6);

			\draw [-> , >=latex ,line width=0.3mm] (-1.5,0) -- (-0.5,0);	
			\node[text width=0.2cm,scale=1.3] at(-0.6,0.3){$x$};
			\draw [-> , >=latex ,line width=0.3mm] (-1.5,0) -- (-1.5,1.);
			\node[text width=0.2cm,scale=1.3] at(-1.2,0.9){$y$};
			\draw node[circle,draw,thick,fill=black, scale=0.2] at (-1.5,0.)  {};
			\draw node[circle,draw, scale=0.3,fill=black] at (3,3) {};

			\node[text width=2cm,scale=1.] at(3.6,-0.5){$x=0m$};
			\node[text width=2cm,scale=1.] at(0.6,-0.5){$x=-1.5m$};
			\node[text width=2cm,scale=1.] at(6.6,-0.5){$x=1.5m$};
			\node[text width=2cm,scale=1.] at(7.5,0.1){$y=-1.5m$};
			\node[text width=2cm,scale=1.] at(7.5,3){$y=0m$};
			\node[text width=2cm,scale=1.] at(7.5,6.){$y=1.5m$};
		\end{tikzpicture}
	\end{center}
\end{figure}
\end{minipage}
\hfill \hfill \vrule \hfill \hfill 
\begin{minipage}[l]{0.45\linewidth}
	  \includegraphics[scale=0.48]{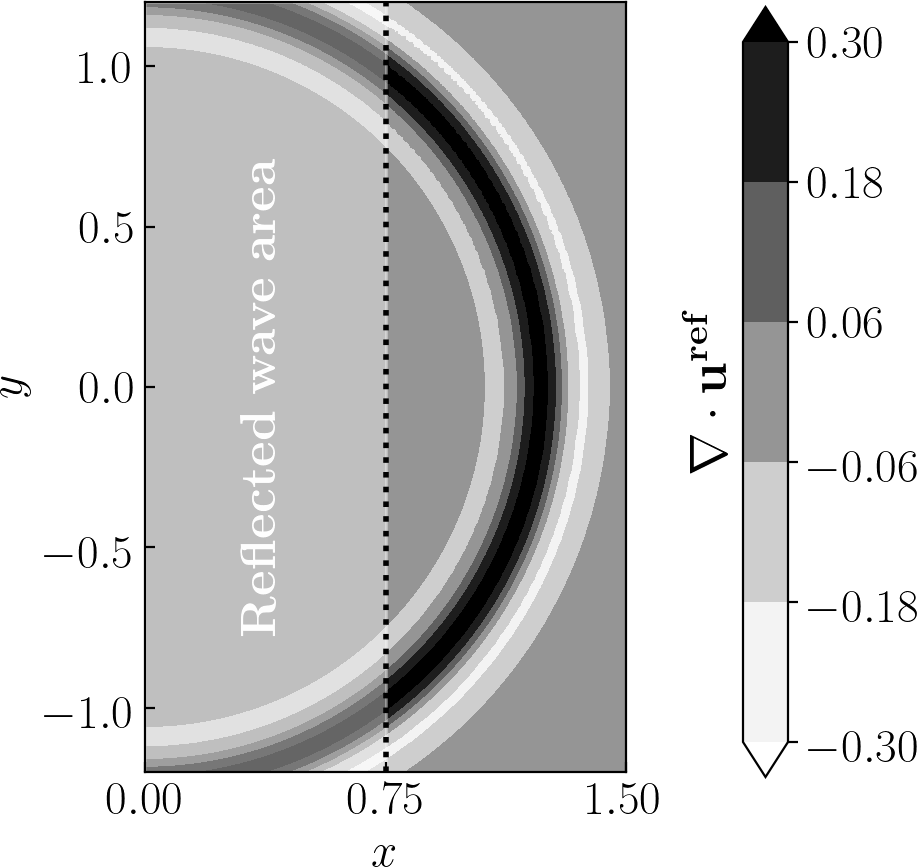}
\end{minipage}
\captionof{figure}{\label{fig:Pulse_domain} Left: Sketch of the simulation domain for the acoustic pulse test case with plane mesh refinement interfaces. Right: velocity divergence field $\left(\mathbf{\nabla \cdot u^{ref}}\right)$ of the pulse with a uniform fine mesh used as a reference for the non-uniform simulations. (\protect\blackdashPulse): refinement interface.\\}

\noindent In this test case, the acoustic reflection induced by the mesh interface is investigated. This spurious reflection can be attributed to three phenomena.

\begin{itemize}
\item A sudden variation in dispersion properties between a fine mesh and a coarse mesh, as evidenced in our last study~\cite{Astoul2020}. This phenomenon is independent of the considered grid refinement algorithm.
\item An aliasing effect. A wave resolved with less than 4 points per wavelength in a fine mesh has no counterpart in a coarse one. A spectral aliasing might thus take place, which results in a reflected acoustic wave. This phenomenon can be attenuated thanks to a filtering step when rescaling information from the fine mesh to the coarse one (Eq.~(\ref{eq:rescff})). However, no noticeable improvement has been obtained with the add of a filtering step on this test case. 
\item The precision of the grid coupling algorithm. A slight discontinuity in the transfer between grids may lead to an acoustic reflection.
\end{itemize}

\begin{figure}[H]
	\begin{center}
		\includegraphics[scale=0.5]{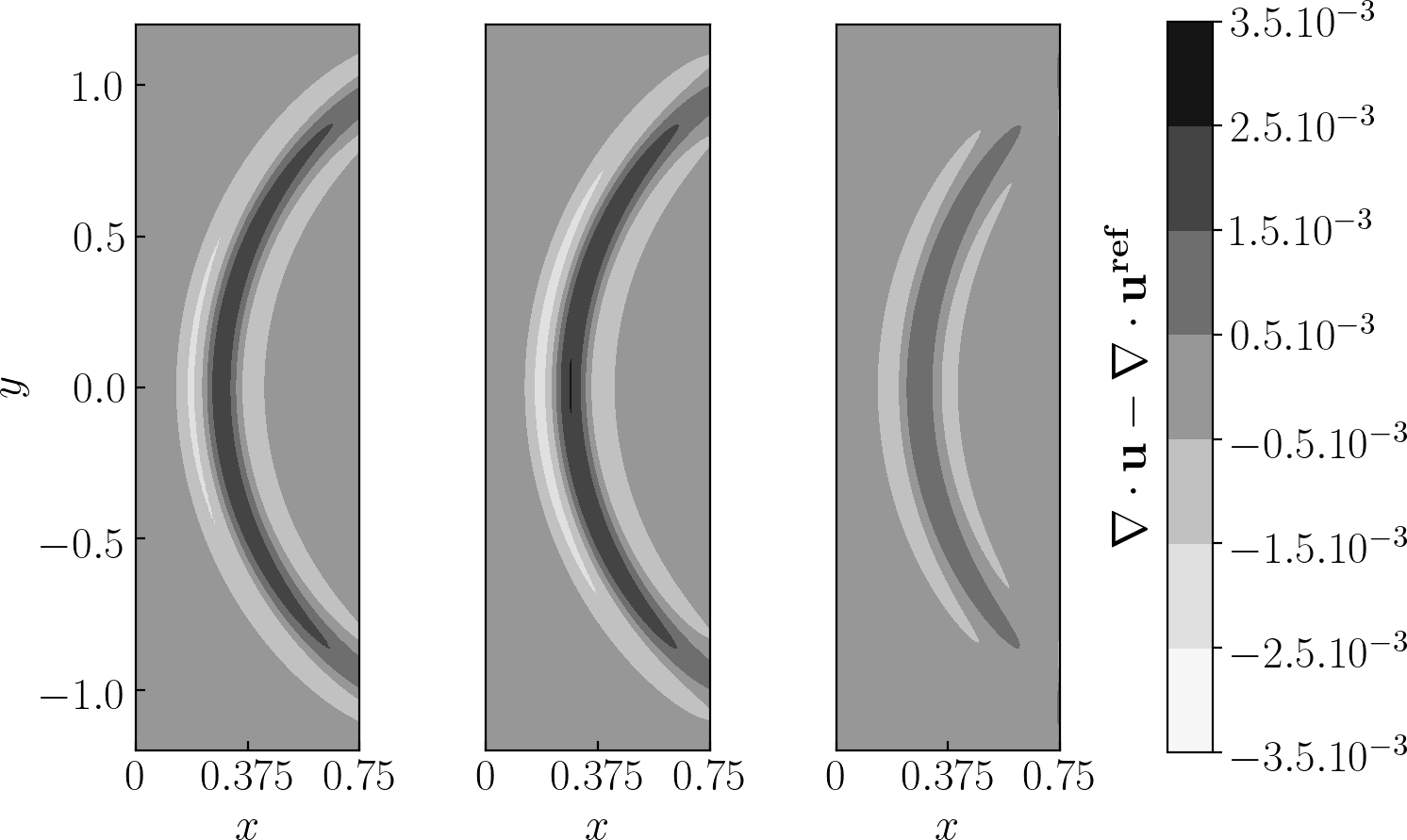}\caption{\label{fig:Pulse_reflexion} Relative velocity divergence field $\left(\mathbf{\nabla \cdot u}-\mathbf{\nabla \cdot u^{ref}}\right)$ of the reflected wave for the pulse test across a plane mesh refinement. The uniform fine simulation is taken as reference $\left(\mathbf{\nabla \cdot u^{ref}}\right)$. Left: STD, middle: DC1, right: DC2.}
	\end{center}
\end{figure}

\noindent Reflection rates in the velocity divergence, obtained with the three grid refinement algorithms (STD, DC1 and DC2), are compared on Fig.~\ref{fig:Pulse_reflexion} for this test case. The uniform simulation displayed on Fig.~\ref{fig:Pulse_domain} is considered as a reference.\\

In any case, an acoustic reflection can be highlighted in non-uniform simulations, due to a combination of the three reasons stated above. With the DC2 algorithm, the amplitude of the spurious reflected wave is significantly reduced compared to that obtained with the DC1 algorithm. This seems to be caused by the fact that in the DC1 algorithm, the reconstruction of the distribution functions is carried out keeping as many fine distribution functions as possible. This leads to a higher aliasing effect, inducing a larger acoustic reflection.

\subsubsection{Acoustic pulse across a circular refinement interface}
\label{subsubsec:pulse_circular}

In this section, a $1m$-radius circular transition is located around the initial position of the acoustic pulse. This type of transition is chosen since it makes it possible to study a wide variety of interface shapes. A sketch of the computational domain is diplayed on Fig.~\ref{fig:PulseCir_mesh}.

\begin{minipage}[l]{0.45\linewidth}
\begin{figure}[H]
	\begin{center}
		\begin{tikzpicture}[scale=0.7]
			\draw (0.,0) grid[step=0.3] (6,6);

			\draw (0.9,3) grid[step=0.15] (5.1,3.3);
			\draw (0.9,3.3) grid[step=0.15] (5.1,3.6);
			\draw (1.2,3.6) grid[step=0.15] (4.8,3.9);
			\draw (1.2,3.9) grid[step=0.15] (4.8,4.2);
			\draw (1.5,4.2) grid[step=0.15] (4.5,4.5);
			\draw (1.8,4.5) grid[step=0.15] (4.2,4.8);
			\draw (2.4,4.8) grid[step=0.15] (3.6,5.1);

			\draw (0.9,2.7) grid[step=0.15] (5.1,3.);
			\draw (0.9,2.7) grid[step=0.15] (5.1,2.4);
			\draw (1.2,2.4) grid[step=0.15] (4.8,2.1);
			\draw (1.2,2.1) grid[step=0.15] (4.8,1.8);
			\draw (1.5,1.8) grid[step=0.15] (4.5,1.5);
			\draw (1.8,1.2) grid[step=0.15] (4.2,1.5);
			\draw (2.4,0.9) grid[step=0.15] (3.6,1.2);

			\draw [-> , >=latex ,line width=0.3mm] (-1.5,0) -- (-0.5,0);	
			\node[text width=0.2cm,scale=1.3] at(-0.6,0.3){$x$};
			\draw [-> , >=latex ,line width=0.3mm] (-1.5,0) -- (-1.5,1.);
			\node[text width=0.2cm,scale=1.3] at(-1.2,0.9){$y$};
			\draw node[circle,draw,thick,fill=black, scale=0.2] at (-1.5,0.)  {};
			\draw node[circle,draw, scale=0.3,fill=black] at (3,3) {};
			\draw node[circle,draw, scale=9,thick,dashed,] at (3,3) {};

			\node[text width=2cm,scale=1.] at(3.6,-0.5){$x=0m$};
			\node[text width=2cm,scale=1.] at(0.6,-0.5){$x=-1.5m$};
			\node[text width=2cm,scale=1.] at(6.6,-0.5){$x=1.5m$};
			\node[text width=2cm,scale=1.] at(7.5,0.1){$y=-1.5m$};
			\node[text width=2cm,scale=1.] at(7.5,3){$y=0m$};
			\node[text width=2cm,scale=1.] at(7.5,6.){$y=1.5m$};
			
			\draw [<-> , >=latex ,line width=0.3mm] (3.,3.) -- (5.15,3.);
			\draw  [fill=white,opacity=1] (3.6,3.15) rectangle (4.5,3.75) ;	
			\node[text width=0.2cm,scale=1.] at(3.9,3.47){$1m$};		
		\end{tikzpicture}
	\end{center}
\end{figure}
\end{minipage}
\hfill \hfill \vrule \hfill \hfill 
\begin{minipage}[l]{0.45\linewidth}
	  \includegraphics[scale=0.48]{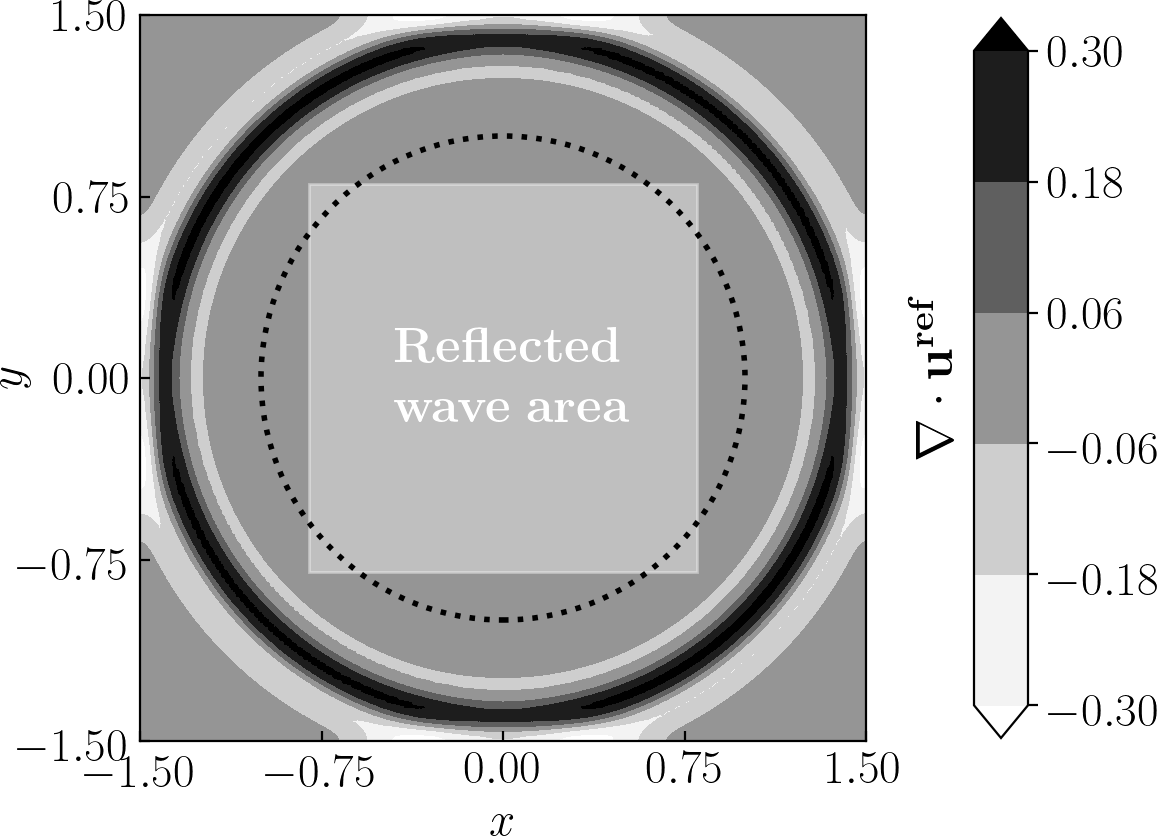}
\end{minipage}
\captionof{figure}{\label{fig:PulseCir_mesh} Left: sketch of the simulation domain for the acoustic pulse test case with a circular mesh refinement interface. Right: velocity divergence field $\left(\mathbf{\nabla \cdot u^{ref}}\right)$ of the pulse with a uniform fine mesh used as reference for the non-uniform simulations. (\protect\blackdashPulse): refinement interface.\\}

\noindent This test case is relevant since it can evidence the isotropy of the acoustic reflection, especially that induced by the spatial interpolation. Even though fourth-order schemes are used regardless the shape of the interface, these interpolations remain one-dimensional. Thus, two non-coincident nodes, although very close to each other, can use interpolation stencils with different normal directions. Therefore, very different interpolation nodes can be used and lead to an anisotropy of the acoustic reflection.\\

\begin{figure}[H]
	\begin{center}
		\includegraphics[scale=0.65]{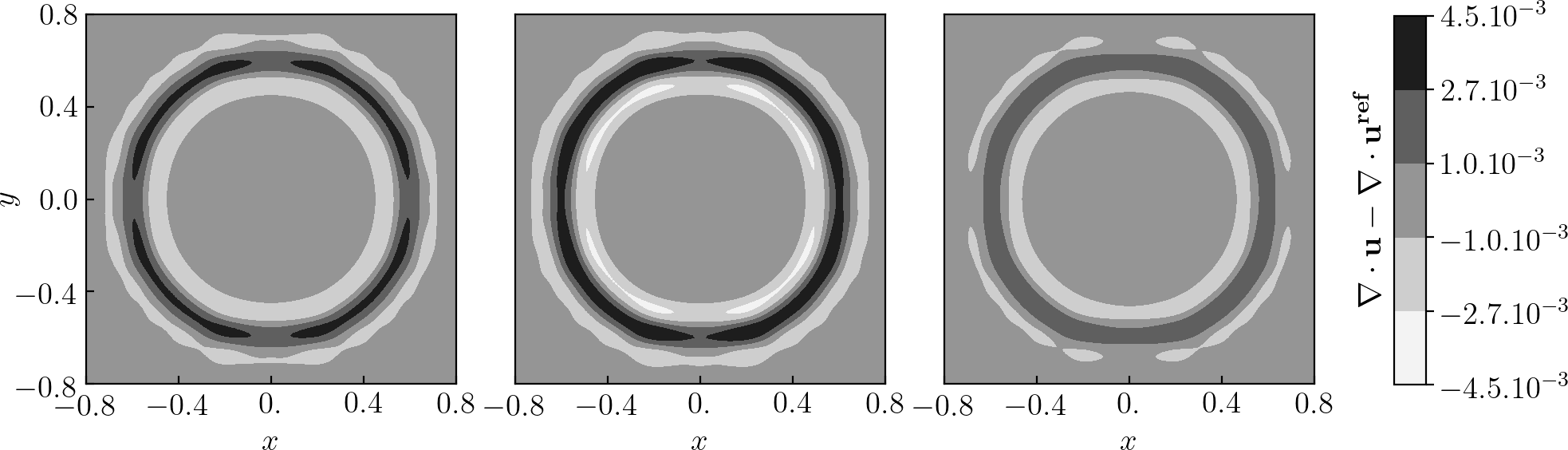}\caption{\label{fig:Pulse_Circ_reflexion} Relative velocity divergence field $\left(\mathbf{\nabla \cdot u}-\mathbf{\nabla \cdot u^{ref}}\right)$ of the reflected wave for the pulse test case across a cylindrical mesh refinement. The uniform fine simulation is taken as reference $\left(\mathbf{\nabla \cdot u^{ref}}\right)$. Left: STD, middle: DC1, right: DC2.}
	\end{center}
\end{figure}

\noindent Here, the acoustic pulse will expand and cross the interface. The reflected spurious wave is shown on Fig.~\ref{fig:Pulse_Circ_reflexion}. The reference used for these simulations is identical to the one used in the previous section since the latter is performed on an uniform mesh~(Fig.~\ref{fig:PulseCir_mesh}).

\noindent The three simulations highlight an anisotropy of the acoustic reflection, which may be caused by the one-dimensional spatial interpolations. The minimal reflection appears along the $x$- and $y$-axes. Along these axes, the cylindrical refinement interface is tangent to the Cartesian mesh, which makes it quasi-planar as schematically displayed on Fig.~\ref{fig:PulseCir_mesh}. The treatment of the non-coincident nodes is therefore optimal because neighboring non-coincident nodes use the same normal direction for interpolations.

\noindent Then, comparing the transition algorithms, it can be observed that the reflected wave with the lowest intensity is produced by the DC2 algorithm whatever the shape of the interface. As previously noticed in the plane transition case, the largest reflection is induced by the DC1 algorithm.\\

\noindent The conclusion retained from these two purely acoustic test cases is that the DC2 algorithm turns out to be the most accurate one for propagating a wave from a fine mesh to a coarse one. The opposite transfer from a coarse to a fine mesh has also been investigated and provides similar conclusions on the algorithm quality.

\subsection{Convected vortex}
\label{subsec:COVO}

The case of a vortex convected across a grid refinement interface is addressed here. It is a standard but very challenging test case for aeroacoustics since orders of magnitude of the aerodynamic pressure fluctuations are several orders of magnitude larger than the pressure variations related to the acoustics. Thus, a very small error in the transmission of the vortex might induce a severe spurious acoustic wave that has to be minimized as much as possible. This test case was used in two previous LBM aeroacoustic studies on non-uniform meshes~\cite{Gendre2017,Astoul2020}.\\

For this test case, a reference to a past article~\cite{Astoul2020} is of paramount importance to get rid of the non-hydrodynamic modes present in the vortex, which can drastically increase the spurious emission. Like in the previous study, these modes are here filtered out by the use of the H-RR collision model.\\

\noindent A barotropic vortex~\cite{Wissocq2020a}, solution of the isothermal Euler equations, is initialized in the fine mesh as follows:

\begin{align}
\label{eq:init_COVO}
		& \rho \left(x,y,z \right) = \rho _0 \exp \left[ -\frac{\epsilon ^2}{2 c_s ^2} \exp\left( -\frac{(x-x_c)^2 + (y-y_c)^2}{R_c ^2} \right) \right], \\
		& u_x\left(x,y,z\right) = U_x - \epsilon \left( \frac{y-y_c}{R_c} \right) \exp \left( - \frac{(x-x_c)^2 + (y-y_c)^2}{2 R_c ^2} \right), \\
	 	& u_y\left(x,y,z\right) = \epsilon \left( \frac{x-x_c}{R_c} \right) \exp \left( - \frac{(x-x_c)^2 + (y-y_c)^2}{2 R_c ^2} \right),  \\
	 	& u_z\left(x,y,z\right) = 0,
\end{align}

\noindent with
\begin{equation}
	\label{eq:init_COVO2}
	\quad \rho_0 = 1\ kg/m^3,
	\quad U_x = 0.1 c_0,
	\quad \epsilon = 0.15U_x,
	\quad \Delta_x^f = 0.01\ m,
    \quad R_c = 0.06\ m,
    \quad (x_c,y_c) = (-6R_c,0).  
\end{equation}

\noindent The vortex convection across a vertical interface is first studied, then a $30^\circ$-inclined interface, which combines many local shapes for the grid interface, will be of interest. It is worth noting that, in all the results presented below, increasing the order of interpolations up to the sixth-order by fetching six neighbors, has no significant effect on the solution, neither for a plane nor for an oblique transition. Thus fourth-order interpolation schemes are kept in the following. \\

\subsubsection{Vortex convection across a vertical refinement interface}
\label{subsubsec:COVO_plane}

\begin{figure}[H]
	\begin{center}
		\begin{tikzpicture}[scale=0.65]
			\draw (0,0) grid[step=0.15] (5.1,6.001);
			\draw (5.1,0) grid[step=0.3] (10.2,6);
			\fill[gray,opacity=0.1] (-0.25,-0.25) rectangle (10.45,0.25) ;
			\draw [dashed] (-0.25,-0.25) rectangle (10.45,0.25) ;
			\fill[gray,opacity=0.1] (-0.25,5.75) rectangle (10.45,6.25) ;
			\draw [dashed] (-0.25,5.75) rectangle (10.45,6.25) ;
			\fill[gray,opacity=0.1] (-0.25,0.25) rectangle (0.25,5.75) ;
			\draw [dashed] (-0.25,0.25) rectangle (0.25,5.75) ;
			\fill[gray,opacity=0.1] (9.95,0.25) rectangle (10.45,5.75) ;
			\draw [dashed] (9.95,0.25) rectangle (10.45,5.75) ;
			\node[text width=3cm,rotate=90,scale=1.4] at(-0.7,3.5){Absorbing Layers};
			\draw  [thick,fill=white,opacity=0.65] (2.5,3) circle (0.8) ;	
			\draw [-> , >=latex ,line width=0.5mm] (2.5,3) -- (4,3);	
			\node[text width=1cm,scale=1.2] at(3.,3.4){$U_0$};
			\draw [-> , >=latex ,line width=0.3mm] (-2.5,0) -- (-1.5,0);	
			\node[text width=0.2cm,scale=1.3] at(-1.6,0.3){$x$};
			\draw [-> , >=latex ,line width=0.3mm] (-2.5,0) -- (-2.5,1.);
			\node[text width=0.2cm,scale=1.3] at(-2.2,0.9){$y$};
			\draw node[circle,draw,thick,fill=black, scale=0.2] at (-2.5,0.)  {};
			\draw node[circle,draw,thick,fill=black, scale=0.3] at (2.5,3.) {};
			\node[text width=1cm,scale=1.] at(2.7,2.7){$x_c,y_c$};
			\node[text width=2cm,scale=1.] at(6.,-0.5){$x=0$};
			\node[text width=2cm,scale=1.] at(0.9,-0.5){$x=-1.5$};
			\node[text width=2cm,scale=1.] at(10.4,-0.5){$x=1.5$};
			\node[text width=2cm,scale=1.] at(12.2,0.1){$y=-1.5$};
			\node[text width=2cm,scale=1.] at(12.2,6.){$y=1.5$};

			\draw [<-,line width=0.3mm,rotate=45](3.9,1.35) arc (90:0:1) ;
		\end{tikzpicture}
		\caption{\label{fig:COVO_mesh} Sketch of the simulation domain for the vortex convected across a plane refinement interface. Absorbing layers map the domain boundaries to avoid reflection of spurious acoustic emission.}
	\end{center}
\end{figure}
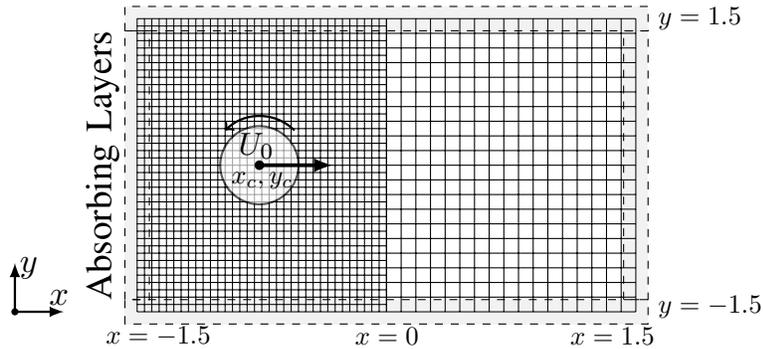

The simulation domain is displayed on Fig.~\ref{fig:COVO_mesh}. A vertical refinement interface is located at $x=0\ m$. The vortex is initialized in the fine grid and is convected from the fine to the coarse mesh. This first case of plane interface allows getting rid of issues related to interpolations and anisotropic treatments of the interface.\\

\noindent In order to avoid any reflection of spurious acoustic, Neumann boundary conditions and explicit absorbing layers are added at the domain boundaries, as previously done in \cite{Chevillotte2016}. In order to map the emitted spurious acoustics, 36 pressure probes are located in a circle at a distance of 1.2m from the domain center.\\

\begin{figure}[H]
	\begin{center}
		\includegraphics[scale=0.47]{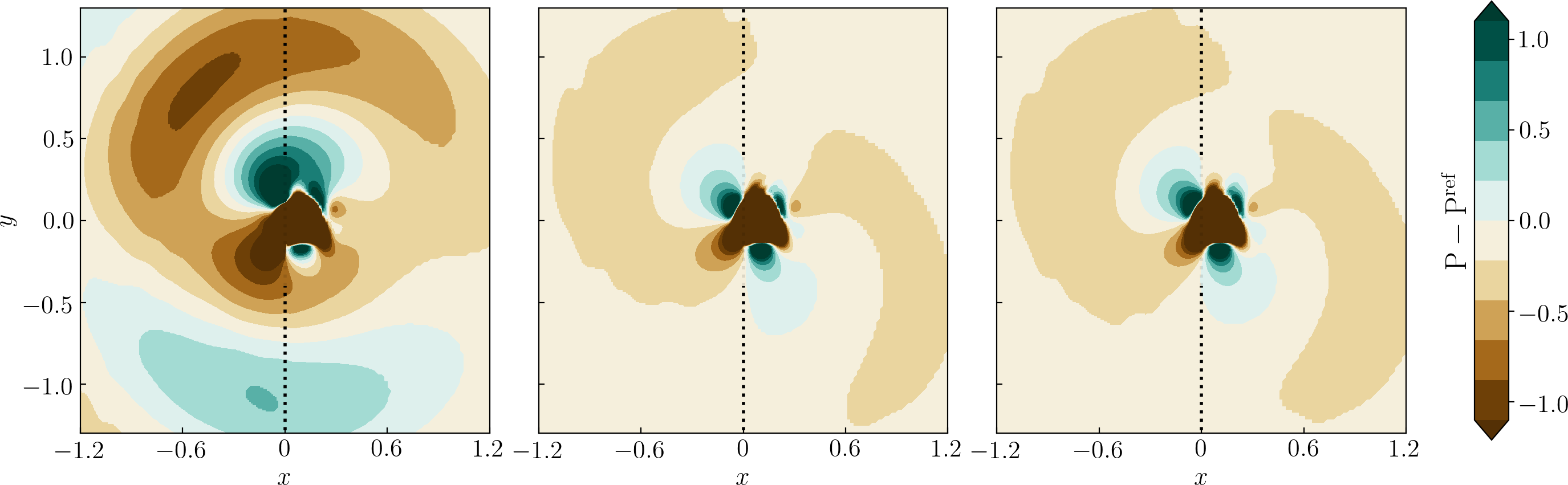}\caption{\label{fig:COVO_plane.png} Relative pressure field $\left(\mathrm{P-P^{ref}}\right)$ of the vortex convected across a plane refinement interface. Left: STD, middle: DC1, right: DC2. (\protect\blackdashPulse): grid refinement interface.}
	\end{center}
\end{figure}

The spurious noise emitted by the vortex is evidenced on Fig.~\ref{fig:COVO_plane.png} by the mean of displaying relative pressure fields. A very strong attenuation of the parasitic noise is obtained with the two $\mathrm{DC}$ algorithms compared to the STD one. This is partly explained by the continuity of the vortex density and velocity, ensured by the $\mathrm{DC}$ algorithms in this case. 

With microphones located in the farfield region, forming an arc around the parasitic source, it is possible to compare OASPLs (Overall Sound Pressure Level)  of spurious noise so as to quantify the intensity and directivity of this emission. There is no reference here, since, theoretically, no acoustic noise is expected by the convection of a single vortex in constant and homogeneous flow. Thus, the whole recorded noise is parasitic. As can be clearly identified on Fig.~\ref{fig:Oaspl_COVO.png}, a large reduction of the parasitic noise is obtained whatever the emitted direction with the $\mathrm{DC}$ algorithms. The azimuthally averaged acoustic emission is then reduced for more than 10dB compared to the STD one. Both  $\mathrm{DC1}$ and $\mathrm{DC2}$ formulations lead to very close emissions, even though the $\mathrm{DC1}$ algorithm turns out to be slightly better on this case (0.9dB of average reduction).

\begin{minipage}[l]{0.45\linewidth}
\begin{figure}[H]
	\includegraphics[scale=0.6]{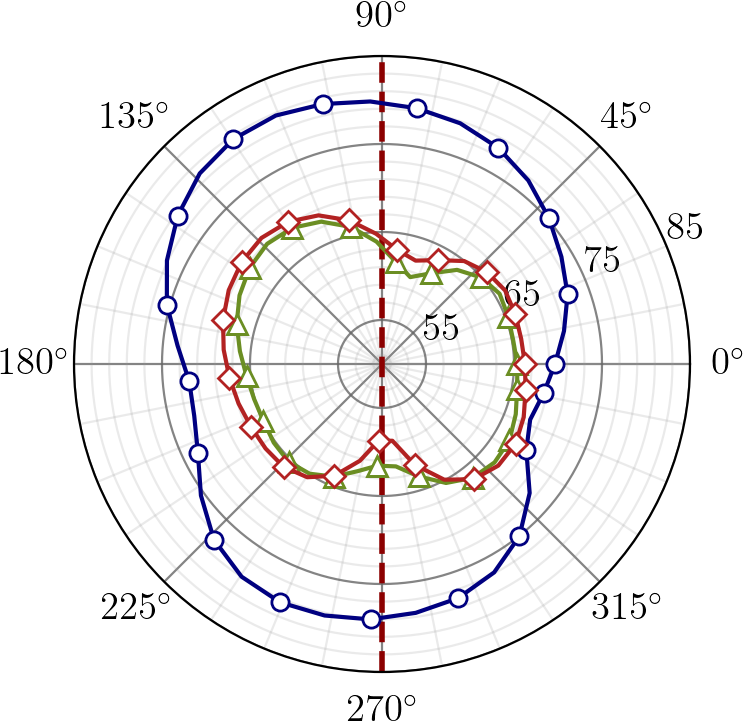}
\end{figure}
\end{minipage}
\hfill \hfill \vrule \hfill \hfill 
\begin{minipage}[l]{0.45\linewidth}
\begin{table}[H]
	\centering
	\begin{tabu}{ c|[1.2pt]c|[1.2pt]c|[1.2pt]c|[1.2pt]}
		 & \textbf{STD} & \textbf{DC1} & \textbf{DC2}\\\tabucline[1.2pt]{-}
		 \textbf{maximum} & \  & \   & \  \\
		 \textbf{OASPL(dB)} &  $\boldsymbol{80.7}$ & $\boldsymbol{68.9}$ & $\boldsymbol{69.8}$\\
 		 \tabucline[1.2pt]{-}
		 \textbf{mean} & \  & \   & \  \\
		 \textbf{OASPL(dB)} &  $\boldsymbol{77.0}$ & $\boldsymbol{65.6}$ & $\boldsymbol{66.5}$	
 		 \\\tabucline[1.2pt]{-}		 
	\end{tabu}
\end{table}

\end{minipage}
\captionof{figure}{\label{fig:Oaspl_COVO.png} Left: pressure OASPL of the spurious acoustics measured in the farfield for the vortex convected across a vertical refinement interface.  \protect\lineSTD:~STD, \protect\lineDCOne:~DC1,  \protect\lineDCTwo:~DC2. Right: table of maximal and average OASPL over the 36 microphones.\\}

The $\mathrm{DC}$ algorithms have proven to be very relevant for the case of the vortex advected across a vertical interface. In the following section, a similar vortex is convected across an oblique refinement interface.

\subsubsection{Vortex convected across an inclined refinement interface}
\label{subsubsec:COVO_inclined}

\begin{figure}[H]
	\begin{center}
		\begin{tikzpicture}[scale=0.65]
			\draw (0,0.0) grid[step=0.15] (3.9,0.3);
			\draw (0,0.3) grid[step=0.15] (4.2,0.6);
			\draw (0,0.6) grid[step=0.15] (4.2,0.9);
			\draw (0,0.9) grid[step=0.15] (4.5,1.2);
			\draw (0,1.2) grid[step=0.15] (4.5,1.5);
			\draw (0,1.5) grid[step=0.15] (4.8,1.8);
			\draw (0,1.8) grid[step=0.15] (4.8,2.1);
			\draw (0,2.1) grid[step=0.15] (5.1,2.4);
			\draw (0,2.4) grid[step=0.15] (5.1,2.7);
			\draw (0,2.7) grid[step=0.15] (5.4,3.0);
			\draw (0,3.0) grid[step=0.15] (5.4,3.3);
			\draw (0,3.3) grid[step=0.15] (5.7,3.6);
			\draw (0,3.6) grid[step=0.15] (5.7,3.9);
			\draw (0,3.9) grid[step=0.15] (6.0,4.2);
			\draw (0,4.2) grid[step=0.15] (6.0,4.5);
			\draw (0,4.5) grid[step=0.15] (6.3,4.8);
			\draw (0,4.8) grid[step=0.15] (6.3,5.1);
			\draw (0,5.1) grid[step=0.15] (6.6,5.4);
			\draw (0,5.4) grid[step=0.15] (6.6,5.7);
			\draw (0,5.7) grid[step=0.15] (6.9,6.0);

			\draw (10.2,0.0) grid[step=0.3] (3.9,0.3);
			\draw (10.2,0.3) grid[step=0.3] (4.2,0.6);
			\draw (10.2,0.6) grid[step=0.3] (4.2,0.9);
			\draw (10.2,0.9) grid[step=0.3] (4.5,1.2);
			\draw (10.2,1.2) grid[step=0.3] (4.5,1.5);
			\draw (10.2,1.5) grid[step=0.3] (4.8,1.8);
			\draw (10.2,1.8) grid[step=0.3] (4.8,2.1);
			\draw (10.2,2.1) grid[step=0.3] (5.1,2.4);
			\draw (10.2,2.4) grid[step=0.3] (5.1,2.7);
			\draw (10.2,2.7) grid[step=0.3] (5.4,3.0);
			\draw (10.2,3.0) grid[step=0.3] (5.4,3.3);
			\draw (10.2,3.3) grid[step=0.3] (5.7,3.6);
			\draw (10.2,3.6) grid[step=0.3] (5.7,3.9);
			\draw (10.2,3.9) grid[step=0.3] (6.0,4.2);
			\draw (10.2,4.2) grid[step=0.3] (6.0,4.5);
			\draw (10.2,4.5) grid[step=0.3] (6.3,4.8);
			\draw (10.2,4.8) grid[step=0.3] (6.3,5.1);
			\draw (10.2,5.1) grid[step=0.3] (6.6,5.4);
			\draw (10.2,5.4) grid[step=0.3] (6.6,5.7);
			\draw (10.2,5.7) grid[step=0.3] (6.9,6.0);
%

			\fill[gray,opacity=0.1] (-0.25,-0.25) rectangle (10.45,0.25) ;
			\draw [dashed] (-0.25,-0.25) rectangle (10.45,0.25) ;
			\fill[gray,opacity=0.1] (-0.25,5.75) rectangle (10.45,6.25) ;
			\draw [dashed] (-0.25,5.75) rectangle (10.45,6.25) ;
			\fill[gray,opacity=0.1] (-0.25,0.25) rectangle (0.25,5.75) ;
			\draw [dashed] (-0.25,0.25) rectangle (0.25,5.75) ;
			\fill[gray,opacity=0.1] (9.95,0.25) rectangle (10.45,5.75) ;
			\draw [dashed] (9.95,0.25) rectangle (10.45,5.75) ;
			\node[text width=3cm,rotate=90,scale=1.4] at(-0.7,3.5){Absorbing Layers};
			\draw  [thick,fill=white,opacity=0.65] (2.5,3) circle (0.8) ;	
			\draw [-> , >=latex ,line width=0.5mm] (2.5,3) -- (4,3);	
			\node[text width=1cm,scale=1.2] at(3.,3.4){$U_0$};
			\draw [-> , >=latex ,line width=0.3mm] (-2.5,0) -- (-1.5,0);	
			\node[text width=0.2cm,scale=1.3] at(-1.6,0.3){$x$};
			\draw [-> , >=latex ,line width=0.3mm] (-2.5,0) -- (-2.5,1.);
			\node[text width=0.2cm,scale=1.3] at(-2.2,0.9){$y$};
			\draw node[circle,draw,thick,fill=black, scale=0.2] at (-2.5,0.)  {};
			\draw node[circle,draw,thick,fill=black, scale=0.3] at (2.5,3.) {};
			\node[text width=1cm,scale=1.] at(2.7,2.7){$x_c,y_c$};
			\node[text width=2cm,scale=1.] at(5.9,-0.5){$x=0$};
			\node[text width=2cm,scale=1.] at(0.9,-0.5){$x=-1.5$};
			\node[text width=2cm,scale=1.] at(10.4,-0.5){$x=1.5$};
			\node[text width=2cm,scale=1.] at(12.2,0.1){$y=-1.5$};
			\node[text width=2cm,scale=1.] at(12.2,6.){$y=1.5$};

			\draw [<-,line width=0.3mm,rotate=45](3.9,1.35) arc (90:0:1) ;
		\end{tikzpicture}
		\caption{\label{fig:COVO_inclined_mesh} Sketch of the simulation domain for the vortex convected across a $30^\circ$-inclined grid interface. Absorbing layers map the domain boundaries to avoid reflection of spurious acoustic emission.}
	\end{center}
\end{figure}
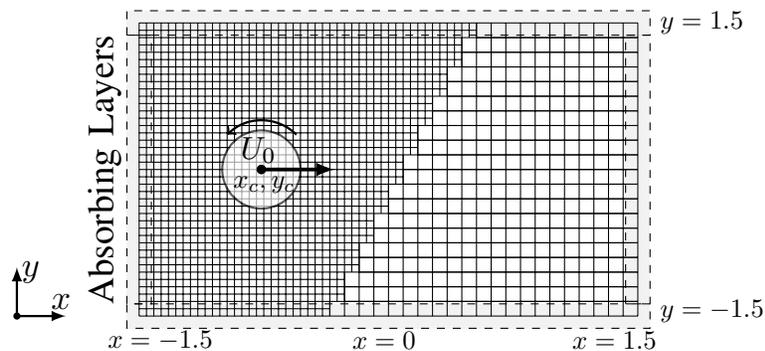

The exactly same data setting as introduced in Sec.~\ref{subsubsec:COVO_plane} is reproduced in this section. However, a $30^\circ$-inclined interface is now considered, as shown on Fig.~\ref{fig:COVO_inclined_mesh}. This case allows assessing the accuracy of the different grid refinement algorithms whatever the local shape of the transition (planar or stepped configuration). This time, spatial interpolations are expected to play a major role. \\

\begin{figure}[H]
	\begin{center}
		\includegraphics[scale=0.47]{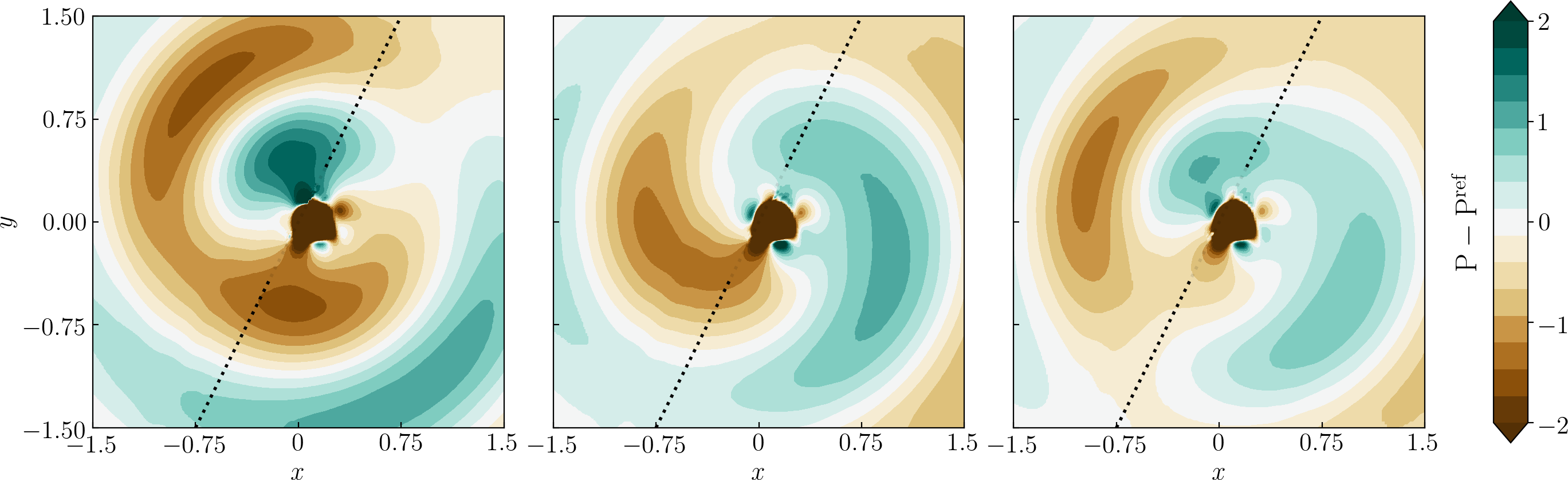}\caption{\label{fig:COVO_inclined.png} Relative pressure field $\left(\mathrm{P-P^{ref}}\right)$ of the vortex convected across an inclined plane refinement interface. Left: STD, middle: DC1, right: DC2. (\protect\blackdashPulse): grid refinement interface.}
	\end{center}
\end{figure}

Spurious acoustics emitted by the vortex convection  across this interface  is displayed on Fig.~\ref{fig:COVO_inclined.png} for the three grid refinement algorithms. 

This time, every algorithm emits spurious acoustics in a similar intensity, and much larger than that obtained with a plane transition, even if better results are, again, obtained with the DC algorithms. In addition, there is a slight discontinuity of the vortex pressure field with the three algorithms, which may be responsible for this emission. This discontinuity can be attributed to the one-dimensional spatial interpolations used, resulting in an anisotropic treatment of non-coincident nodes.

More quantitatively, pressure OASPLs, displayed on Fig.~\ref{fig:Oaspl_COVO_inclined.png}, indicate that the emission of the three algorithms (STD, DC1, DC2) are, indeed, of the same order of magnitude. Averaged OASPLs are well above those obtained for a plane interface with an increase of 6, 15 and 13 dB respectively. Thus, the DC1 algorithm is the most degraded one by the use of oblique transitions. For this one, a very intense wave (up to 84.5dB) is observed in the coarse mesh, whereas the maximal acoustic amplitude produced in the fine mesh is 4dB less intense. This is all the more problematic as for aeroacoustic applications, microphones are generally located in the farfield region, \textit{i.e.} in the direction of the coarse mesh. They would therefore be subject to a more intense parasitic emission. This time, the DC2 algorithm is the most conclusive one on this test case, although the benefit compared to the STD one is reduced on this inclined transition. Furthermore, the emission is better distributed between both grids with the DC2 algorithm,.

\begin{minipage}[l]{0.45\linewidth}
\begin{figure}[H]
	\includegraphics[scale=0.6]{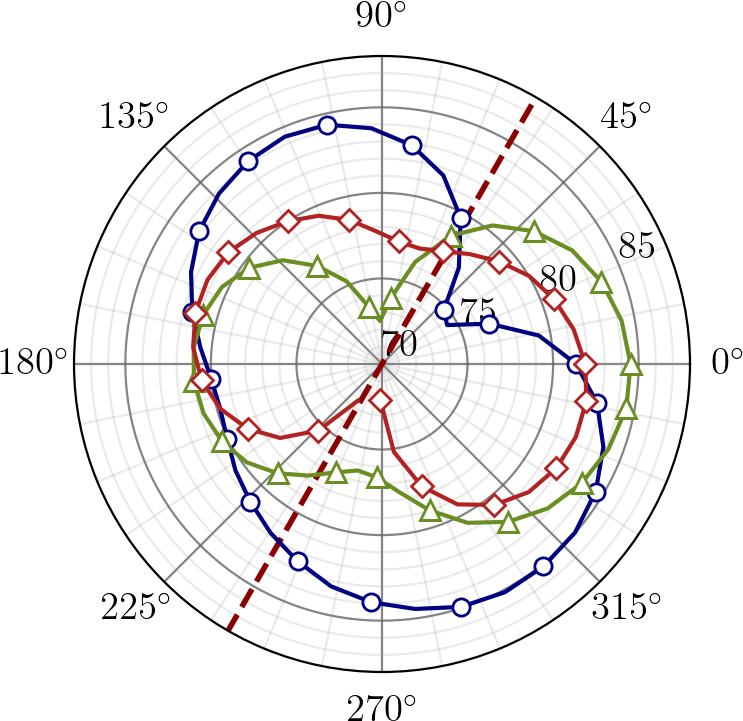}
\end{figure}
\end{minipage}
\hfill \hfill \vrule \hfill \hfill 
\begin{minipage}[l]{0.45\linewidth}
\begin{table}[H]
	\centering
	\begin{tabu}{ c|[1.2pt]c|[1.2pt]c|[1.2pt]c|[1.2pt]}
		 & \textbf{STD} & \textbf{DC1} & \textbf{DC2}\\\tabucline[1.2pt]{-}
		 \textbf{maximum} & \  & \   & \  \\
		 \textbf{OASPL(dB)} &  $\boldsymbol{85.1}$ & $\boldsymbol{84.5}$ & $\boldsymbol{82.1}$\\
		 \tabucline[1.2pt]{-}
		 \textbf{mean} & \  & \   & \  \\
		 \textbf{OASPL(dB)} &  $\boldsymbol{82.6}$ & $\boldsymbol{80.9}$ & $\boldsymbol{79.6}$
 		 \\\tabucline[1.2pt]{-}
	\end{tabu}
\end{table}
\end{minipage}
\captionof{figure}{\label{fig:Oaspl_COVO_inclined.png} Left: Pressure OASPL of the spurious acoustic measured in the farfield for the convected vortex test case that crosses an inclined plane refinement interface. \protect\lineSTD:~STD, \protect\lineDCOne:~DC1,  \protect\lineDCTwo:~DC2. Right: Table of maximum and average OASPL over the 36 microphones. (\protect\redBrickdashed): grid refinement interface.\\}

From these academic test cases, the following conclusions can be drawn whatever the shape of the transition.
\begin{itemize}
	\item The $\mathrm{DC2}$ algorithm is the most relevant one to deal with pure acoustics.
	\item $\mathrm{DC}$ algorithms are much more accurate than the STD one to convect a vortex from a given mesh resolution to another. However, the benefits of these algorithms are strongly degraded by the quality of interpolations in the presence of arbitrary inclined transition. In that case, a slightly better behavior has been observed with the $\mathrm{DC2}$ algorithm, especially when focusing on the acoustics propagated towards the coarse mesh. This is more critical for aeroacoustic applications, as microphones are generally positioned in the farfield.
\end{itemize}

The last remark is of paramount importance: when using this type of interpolation, one should keep in mind that vortices convected across inclined interfaces are likely to generate more spurious noise. This specificity must be taken into account in the mesh design for aeroacoustic applications with complex geometries. For example, arbitrary layers of cells can be used close to the walls and plane grid interfaces should be preferred in the wake region, where many intense vortices are expected.\\

However, cubic refinement boxes must also be used with care, to prevent the creation of acoustic resonators. It has been indeed seen that the passage of an acoustic wave across an interface inevitably generates a reflected wave. This wave can be further reflected on a plane transition placed at the opposite of the refinement box resulting in a stationary wave that could not be evacuated.\\

In the next section, algorithm comparisons will be performed on a turbulent case of a cylinder at high Reynolds number, under typical conditions of an industrial aeroacoustic application.


\section{Numerical validation and comparison with an existing grid refinement algorithm on a turbulent test case}
\label{sec:valid_turbulent}

The purpose of this section is to evaluate the accuracy and stability of the grid refinement algorithms with arbitrary transitions in the presence of turbulent flows. A turbulent cylinder wake is chosen with a Reynolds number set in the super-critical flow regime \cite{Achenbach1971,Achenbach1975}, that is representative of typical aeronautical applications. A comprehensive study of the flow physics will not be performed here, since it mainly depends on the parietal modeling, which is not the subject of this paper. The aim is to simulate a low-viscosity turbulent flow across refinement interfaces minimizing the generation of spurious noise.\\

\noindent The ability of the H-RR collision model to eliminate parasitic vorticity, which is likely to appear at mesh transitions with other collision models, has been shown in a previous work~\cite{Astoul2020}. Comparisons are therefore carried out with a flow that is free from spurious vorticity in the fluid core. Moreover, mesh refinement interfaces will be located far away from the cylinder so that their influence on the wake physics can be considered as weak.
Hence, the global noise emitted in this simulation is a superposition of the dipole noise emitted by the cylinder, and the parasitic noise due to mesh transitions. 
Considering that the dipole noise sources are located at the cylinder wall, and considering that the transitions are far from the latter boundary condition, it can be inferred that physical noise sources are identical whatever the transition algorithm used. Consequently, any additional noise is considered as spurious and will be quantified in this study as such.\\

\noindent In this section, the test case is firstly declined using box-shaped transitions to minimize interpolation errors. In a second step, mesh layers surrounding the cylinder are considered instead of the first cubic box resolution. This second case makes it more representative of industrial meshes where one would like to refine boundary layers.

\subsection{Simulation of the turbulent cylinder with box-shaped grid interfaces}
\label{subsec:cylinder_box}

\begin{figure}[H]
	\begin{center}
		\begin{tikzpicture}[scale=0.6]
			\draw (2.4,2.4) grid[step=0.15] (6.,4.8);
			\draw (1.8,1.8) grid[step=0.3] (9.6,5.4);
			\draw (0.,0) grid[step=0.6] (12.6,7.2);

			\fill[gray,opacity=0.1] (-0.25,-0.25) rectangle (12.85,0.25) ;
			\draw [dashed] (-0.25,-0.25) rectangle (12.85,0.25) ;
			\fill[gray,opacity=0.1] (-0.25,6.9) rectangle (12.85,7.45) ;
			\draw [dashed] (-0.25,6.9) rectangle (12.85,7.45) ;
			\fill[gray,opacity=0.5] (-0.25,0.25) rectangle (0.25,6.9) ;
			\draw [dashed] (-0.25,0.25) rectangle (0.25,6.9) ;
			\fill[gray,opacity=0.1] (12.35,0.25) rectangle (12.85,7.45) ;
			\draw [dashed] (12.35,0.25) rectangle (12.85,7.45) ;
			\node[text width=4.5cm,rotate=90,scale=1.,darkgray] at(-0.65,4.){Absorbing Layers - Velocity};
			\node[text width=4.5cm,rotate=-90,scale=1.,gray] at(13.28,3.2){Absorbing Layers - Pressure};

			\draw  [thick,fill=white,opacity=1] (3.5,3.6) circle (0.8) ;	
			\draw [<->] (2.75,3.6) -- (3.5,3.6) -- (4.25,3.6);
			\node[text width=0cm,rotate=0,scale=0.8] at(3.35,3.95){D};

			\draw  [fill=white,opacity=1] (4.5,4.2) rectangle (6.,4.8) ;	
			\draw  [fill=white,opacity=1] (8.1,4.8) rectangle (9.6,5.4) ;	
			\draw  [fill=white,opacity=1] (10.2,6.) rectangle (11.4,6.6) ;	

			\node[text width=0cm,rotate=0,scale=1.] at(4.7,4.5){RD1};
			\node[text width=0cm,rotate=0,scale=1.] at(8.3,5.1){RD2};
			\node[text width=0cm,rotate=0,scale=1.] at(10.23,6.3){RD3};


			\draw [-> , >=latex ,line width=0.3mm] (-2.5,3.6) -- (-1.5,3.6);	
			\node[text width=0.2cm,scale=1.3] at(-1.6,4.){$y$};
			\draw [-> , >=latex ,line width=0.3mm] (-2.5,3.6) -- (-2.5,2.6);
			\node[text width=0.2cm,scale=1.3] at(-2.1,2.6){$x$};
			\draw node[circle,draw,thick,fill=black, scale=0.2] at (-2.5,3.6)  {};
		\end{tikzpicture}
		\caption{\label{fig:cylinder_box_domain} Sketch of the simulation domain for the cylinder test case. Three refinement domains (RD) are used. Absorbing layers map the domain boundaries to avoid any acoustic reflection.}
	\end{center}
\end{figure}
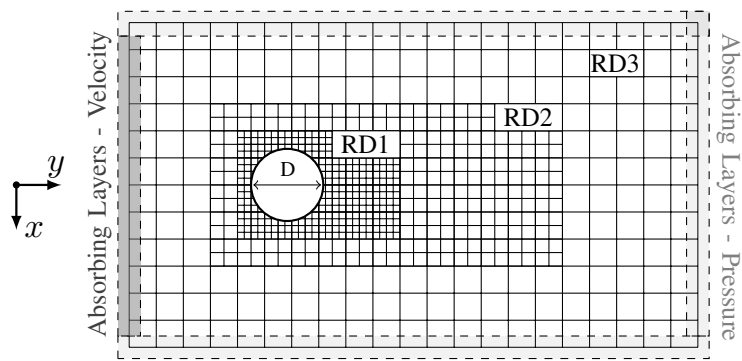

A sketch of the simulation domain is shown on Fig.~\ref{fig:cylinder_box_domain}. Three resolution domains (RD) are placed around the cylinder, on which a wall law taking into account adverse pressure gradients and curvature effects is applied \cite{Afzal1999,Wilhelm2018}. These boxes are placed in such a way that the mesh in the wake is fine enough to ensure the development of turbulent structures before crossing the interfaces. 36 Probes are placed on a 1.5m-radius circle centered  around the cylinder to record acoustic directivity.\\

The simulation setups are

\begin{equation}
	\label{eq:cylinder}
	\quad \mathrm{M}_\infty= 0.1,
	\quad \rho_\infty = 1\ kg.m^{-3},
	\quad \mathrm{\Delta x^f} = 0.001m,
    \quad \mathrm{D} = 0.3m,
    \quad \nu = 1.49.10^{-5}m^{2}.s^{-1},
    \quad \mathrm{T} = 0.5s,
\end{equation}
where $\mathrm{M}_\infty$ is the free stream Mach number imposed at the inlet, $\mathrm{D}$ is the diameter of the cylinder, $\nu$ is the kinematic viscosity and $\mathrm{T}$ the overall simulation time.

A Dirichlet velocity boundary condition is imposed at the inlet and a Dirichlet density boundary condition at the outlets. Both conditions are implemented using a full reconstruction of distribution functions estimated with finite differences as in \cite{Latt2008,Verschaeve2010}. A thickness of $1.6\mathrm{D}$ is chosen in the third dimension ($z$ axis) in order to allow the three-dimensional turbulence to be fully developed. Furthermore, absorbing layers \cite{Chevillotte2016} map the domain boundaries to avoid acoustic reflections and reduce the spurious noise that may be caused by the impact of the turbulent wake on the outlet Dirichlet condition.

\begin{figure}[H]
	\begin{center}
		\includegraphics[scale=0.47]{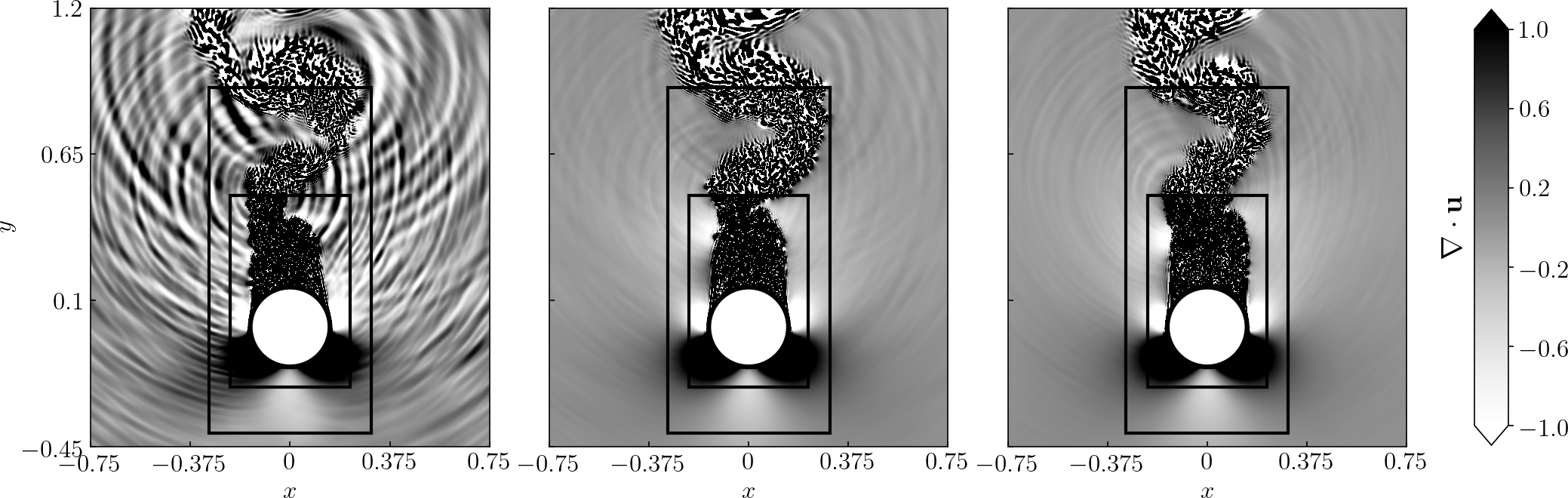}\caption{\label{fig:cylinder_boxes_Dilatation.png} Velocity dilatation field $\left(\mathbf{\nabla \cdot u}\right)$ of the turbulent flow around a cylinder with box-shaped grid interfaces. Left: STD, middle: DC1, right: DC2.}
	\end{center}
\end{figure}

Velocity dilatation fields are displayed on Fig.~\ref{fig:cylinder_boxes_Dilatation.png}, where colormaps are are tightened to highlight acoustic wave fronts. A very important decrease in the spurious noise can be observed with the $\mathrm{DC}$ algorithms, as expected from the results obtained with the convected vortex of Sec.~\ref{subsec:COVO}. Both sources, from RD1 and RD2, are highly attenuated. Finally, it can be noted that no harmful numerical artifacts or parasitic vorticity can be observed at the transitions despite this very tight colormap, whatever the algorithm.

Pressure OASPLs are displayed on Fig.~\ref{fig:Oaspl_cylinder_boxes.png}. A significant reduction of the spurious noise, of about 20dB, is obtained between the STD and the $\mathrm{DC}$ algorithms, whatever the directivity. The DC2 algorithm is the one generating fewest parasitic acoustics (on average 2.2dB less than the DC1 algorithm). These results were not expected with regards to the test case of the convected vortex crossing a plane interface, where the DC1 algorithm seemed to emit less parasitic acoustics than the DC2 algorithm (0.9 dB less in that case).\\

However, all the simulations performed on turbulent test cases led to the same conclusions: the DC2 algorithm is more accurate in the handling of turbulent flows. This may be explained by the fact that using coarse distribution functions in the reconstruction of $f_i$ acts as a partial filtering step. Fine distribution functions have indeed a richer spectral content thanks to the ability of the fine mesh to resolve smaller structures. If this spectral content is directly transferred to the coarse mesh, an aliasing effect may occur, since high frequency waves of the fine mesh does not have any counterpart in the coarse one. Ideally, all fine functions used during the fine to coarse transfer should be filtered \cite{Lagrava2012,Touil2014}. However, any common isotropic filter cannot be used on transition nodes of non-overlapping algorithms, for which some distributions are unknown on the coarse side. Hence, reconstructing distributions with as many coarse distributions as possible minimizes the aliasing phenomenon. This can explain the minimal noise recorded on OASPLs with the DC2 algorithm.

\begin{minipage}[l]{0.45\linewidth}
\begin{figure}[H]
	\includegraphics[scale=0.6]{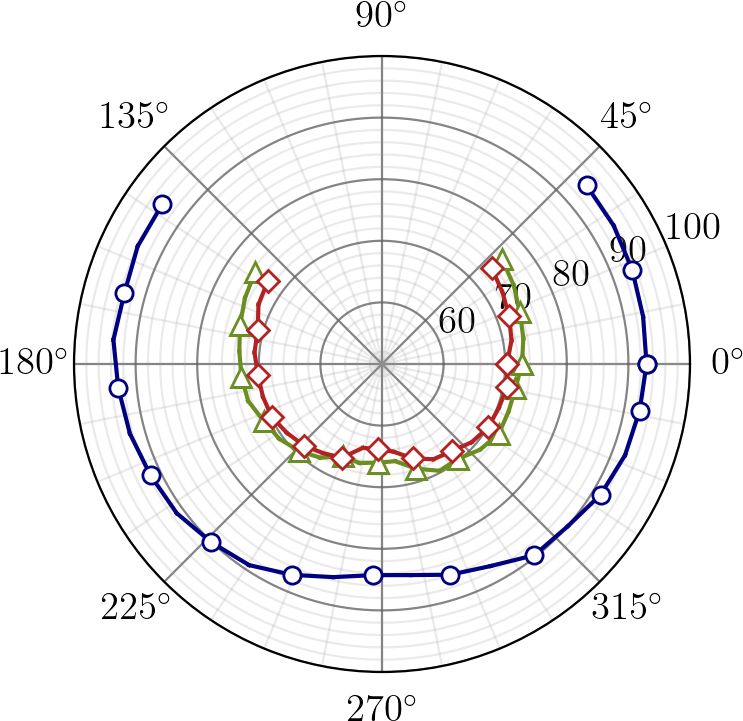}
\end{figure}
\end{minipage}
\hfill \hfill \vrule \hfill \hfill 
\begin{minipage}[l]{0.45\linewidth}
\begin{table}[H]
	\centering
	\begin{tabu}{ c|[1.2pt]c|[1.2pt]c|[1.2pt]c|[1.2pt]}
		 & \textbf{STD} & \textbf{DC1} & \textbf{DC2}\\\tabucline[1.2pt]{-}
		 \textbf{maximal} & \  & \   & \  \\
		 \textbf{OASPL(dB)} &  $\boldsymbol{94.1}$ & $\boldsymbol{75.8}$ & $\boldsymbol{73.7}$
		 \\\tabucline[1.2pt]{-}
		 \textbf{average} & \  & \   & \  \\
		 \textbf{OASPL(dB)} &  $\boldsymbol{91.7}$ & $\boldsymbol{72.2}$ & $\boldsymbol{70.0}$
		 \\\tabucline[1.2pt]{-}
	\end{tabu}
\end{table}
\end{minipage}
\captionof{figure}{\label{fig:Oaspl_cylinder_boxes.png} Left: Pressure OASPL measured in the farfield of the turbulent cylinder with box-shaped grid interfaces. \protect\lineSTD:~STD, \protect\lineDCOne:~DC1,  \protect\lineDCTwo:~DC2. Right: table of maximal and average OASPL over the microphones.\\}

In the following section, additional simulations of the same flow configuration are carried out with a cylindrical RD1 resolution domain.

\subsection{Simulation of the turbulent cylinder with mixed layers and box-shaped grid interfaces\\}
\label{subsec:cylinder_layers}

Offset mesh layers are widely used in industrial simulations, as shown for example in recent LBM studies~\cite{Hou2019,Konig2017,Konig2018}. They make it possible to refine the grid close to the walls, and thus to capture the physics of boundary layers that drive much of the flow physics on realistic geometries.

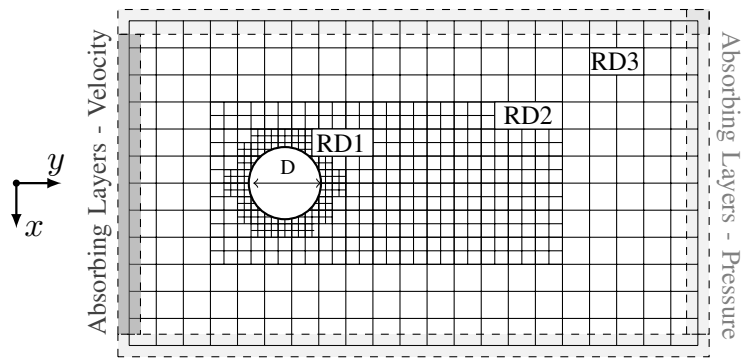
\begin{figure}[H]
	\begin{center}
		\begin{tikzpicture}[scale=0.6]
			\draw (1.8,1.8) grid[step=0.3] (9.6,5.4);
			\draw (0.,0) grid[step=0.6] (12.6,7.2);
			
			\draw (2.7,4.5) grid[step=0.15] (4.1,4.8);
			\draw (2.4,4.2) grid[step=0.15] (4.5,4.5);
			\draw (2.4,3.9) grid[step=0.15] (4.5,4.2);
			\draw (2.1,3.6) grid[step=0.15] (4.8,3.9);
			\draw (2.1,3.3) grid[step=0.15] (4.8,3.6);
			\draw (2.4,3.) grid[step=0.15] (4.5,3.3);
			\draw (2.4,2.7) grid[step=0.15] (4.5,3.);
			\draw (2.7,2.4) grid[step=0.15] (4.1,2.7);

			\fill[gray,opacity=0.1] (-0.25,-0.25) rectangle (12.85,0.25) ;
			\draw [dashed] (-0.25,-0.25) rectangle (12.85,0.25) ;
			\fill[gray,opacity=0.1] (-0.25,6.9) rectangle (12.85,7.45) ;
			\draw [dashed] (-0.25,6.9) rectangle (12.85,7.45) ;
			\fill[gray,opacity=0.5] (-0.25,0.25) rectangle (0.25,6.9) ;
			\draw [dashed] (-0.25,0.25) rectangle (0.25,6.9) ;
			\fill[gray,opacity=0.1] (12.35,0.25) rectangle (12.85,7.45) ;
			\draw [dashed] (12.35,0.25) rectangle (12.85,7.45) ;
			\node[text width=4.5cm,rotate=90,scale=1.,darkgray] at(-0.65,4.){Absorbing Layers - Velocity};
			\node[text width=4.5cm,rotate=-90,scale=1.,gray] at(13.28,3.2){Absorbing Layers - Pressure};

			\draw  [thick,fill=white,opacity=1] (3.45,3.6) circle (0.8) ;	
			\draw [<->] (2.75,3.6) -- (3.5,3.6) -- (4.25,3.6);
			\node[text width=0cm,rotate=0,scale=0.8] at(3.35,3.95){D};
					

			\draw  [fill=white,opacity=1] (4.05,4.2) rectangle (5.4,4.8) ;	
			\draw  [fill=white,opacity=1] (8.1,4.8) rectangle (9.6,5.4) ;	
			\draw  [fill=white,opacity=1] (10.2,6.) rectangle (11.4,6.6) ;	

			\node[text width=0cm,rotate=0,scale=1.] at(4.15,4.5){RD1};
			\node[text width=0cm,rotate=0,scale=1.] at(8.3,5.1){RD2};
			\node[text width=0cm,rotate=0,scale=1.] at(10.23,6.3){RD3};

			\draw [-> , >=latex ,line width=0.3mm] (-2.5,3.6) -- (-1.5,3.6);	
			\node[text width=0.2cm,scale=1.3] at(-1.6,4.){$y$};
			\draw [-> , >=latex ,line width=0.3mm] (-2.5,3.6) -- (-2.5,2.6);
			\node[text width=0.2cm,scale=1.3] at(-2.1,2.6){$x$};
			\draw node[circle,draw,thick,fill=black, scale=0.2] at (-2.5,3.6)  {};
		\end{tikzpicture}
		\caption{\label{fig:cylinder_layers_domain.png} Sketch of the simulation domain for the cylinder test case. Three refinement domains (RD) are used, the first one being cylindrical. Absorbing layers map the domain boundaries to avoid reflection of spurious acoustic emission.}
	\end{center}
\end{figure}

The adopted mesh is shown on Fig.~\ref{fig:cylinder_layers_domain.png}, where a layer of 75 cells is considered in RD1. This very large distance between the first refinement interface and the solid wall allows better visualizing the flow and the numerical artifacts that might occur. This choice is not adopted for a physically optimized simulation, where generally layers of 6 to 7 cells are designed close to the walls.

Velocity divergence fields are shown on Fig.~\ref{fig:cylinder_boxes_Dilatation.png}. Like in the previous section, a very strong reduction of the parasitic noise is observed with the $\mathrm{DC}$ algorithms. No numerical artifacts, except parasitic acoustics, appear on the cylindrical transitions whatever the algorithm. The acoustic source generated by the RD1 one is less intense than that observed on Fig.~\ref{fig:cylinder_boxes_Dilatation.png}. This may be due to the fact that the transition is closer to the cylinder. Turbulent structures that are convected across it are then smaller and less intense. These structures therefore produce less acoustic noise, even though the quality of the algorithm may be degraded by spatial interpolations as discussed in Sec.~\ref{sec:valid_academic}.

\begin{figure}[H]
	\begin{center}
		\includegraphics[scale=0.47]{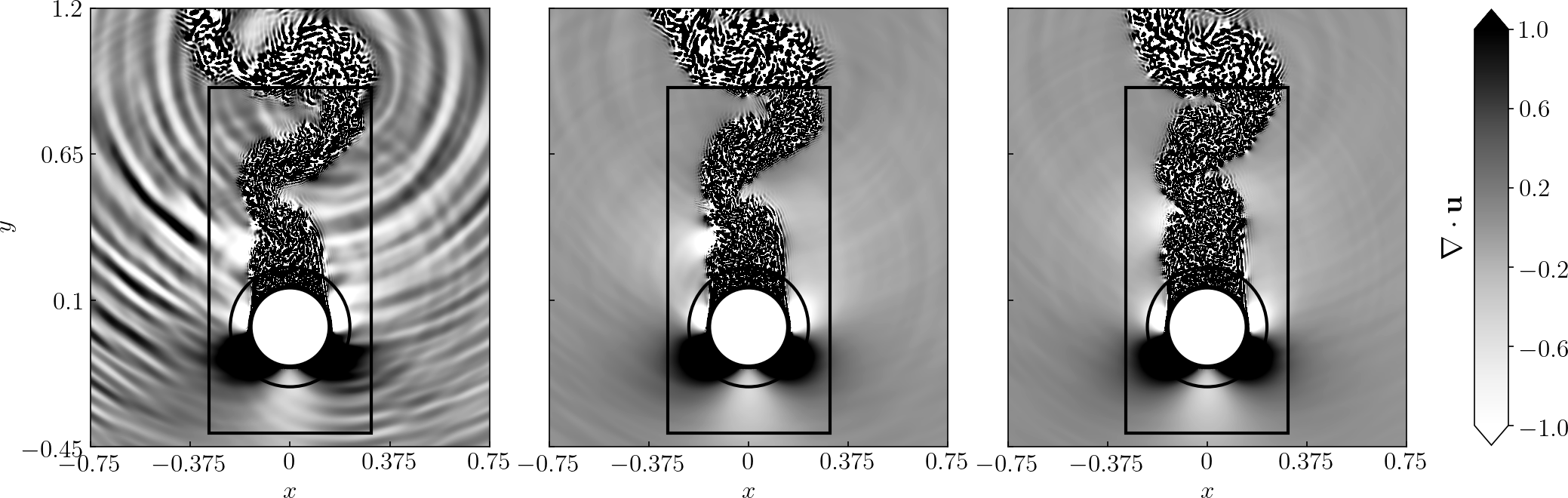}\caption{\label{fig:cylinder_layers_Dilatation.png} Velocity dilatation field $\left(\mathbf{\nabla \cdot u}\right)$ of the turbulent cylinder with a mixed of layers and box-shaped grid interface. Left: STD, middle: DC1, right: DC2.}
	\end{center}
\end{figure}

More quantitatively, it can be seen on the OASPLs (Fig.~\ref{fig:Oaspl_cylinder_layers.png}) that a significant decrease in spurious noise is again observed with the $\mathrm{DC}$ algorithms, with notably 21dB reduction between STD and DC2 algorithms. The simulation with the DC1 algorithm is the only one where the average noise has been increased by the use of cylindrical transitions. The difference with the DC2 algorithm is thus increased from 2.2dB to 2.6dB. This result was expected, since Sec.~\ref{subsubsec:COVO_inclined} highlighted that the DC1 algorithm could emit significantly more noise to the far field than the DC2 one with a non-planar interface.

\begin{minipage}[l]{0.45\linewidth}
\begin{figure}[H]
	\includegraphics[scale=0.6]{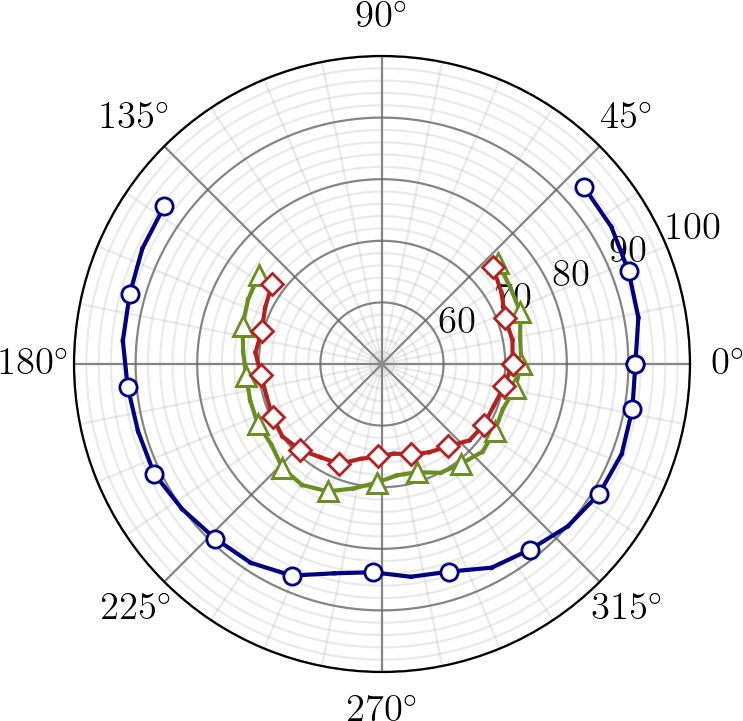}
\end{figure}
\end{minipage}
\hfill \hfill \vrule \hfill \hfill 
\begin{minipage}[l]{0.45\linewidth}
\begin{table}[H]
	\centering
	\begin{tabu}{ c|[1.2pt]c|[1.2pt]c|[1.2pt]c|[1.2pt]}
		 & \textbf{STD} & \textbf{DC1} & \textbf{DC2}\\\tabucline[1.2pt]{-}
		 \textbf{maximal} & \  & \   & \  \\
		 \textbf{OASPL(dB)} &  $\boldsymbol{93.7}$ & $\boldsymbol{75.0}$ & $\boldsymbol{73.9}$
		 \\\tabucline[1.2pt]{-}
		 \textbf{average} & \  & \   & \  \\
		 \textbf{OASPL(dB)} &  $\boldsymbol{90.9}$ & $\boldsymbol{72.5}$ & $\boldsymbol{69.9}$
		 \\\tabucline[1.2pt]{-}
	\end{tabu}
\end{table}
\end{minipage}
\captionof{figure}{\label{fig:Oaspl_cylinder_layers.png} Left: Pressure OASPL measured in the farfield of the turbulent cylinder with a combination of cylindrical and box-shaped grid interfaces. \protect\lineSTD:~STD, \protect\lineDCOne:~DC1,  \protect\lineDCTwo:~DC2. Right: Table of maximal and average OASPL over the microphones.\\}

These turbulent test cases allow for validating the $\mathrm{DC}$ algorithms under realistic conditions, typical of industrial aeroacoustic applications. These algorithms do not present any stability issue and offer a very important gain in accuracy. They are key elements in addition to the H-RR collision model (or any model filtering out non hydrodynamic modes) to perform aeroacoustic simulations that are not polluted by parasitic noise. More specifically, the DC2 algorithm is more relevant for dealing with turbulent flows than the DC1 algorithm. This might be attributed to the use of coarse distribution functions in the reconstruction, which reduces aliasing effects. Furthermore, since this formulation is more accurate when using transitions of any shape, as well as to transmit acoustic waves, the DC2 algorithm should be preferred for aeroacoustic simulations.


\section{Conclusion}
\label{sec:conclusion}

In this study, a new family of algorithms has been proposed to cope with the abrupt resolution transitions of Lattice-Boltzmann non-uniform grids. Based on a direct grid coupling formulation, these types of algorithms distinguish from commonly used methods, where overlapped grid areas are generally considered. The proposed algorithms are based on the use of a unique and consistent equilibrium distribution function at the mesh interface, which complies with the conservation of mass and momentum. This equilibrium function has been used to reconstruct the missing distribution functions at both coarse and fine mesh sides, as well as in the collision step. The way missing distributions are reconstructed is not unique since some of them are known in both meshes: one can either conserve fine or coarse populations. This choice has a significant influence on the accuracy of the algorithm. In this study, two possible reconstructions have been addressed and compared. Validations and comparisons on academic cases have been carried out for arbitrary transitions. First of all,  the DC2 algorithm proved to be the most accurate one for a purely acoustic pulse crossing a mesh transition. Then, a vortex convected across a transition has been studied. For this case, the spurious noise generated by the DC1 algorithm was slightly lower than that generated by the DC2 algorithm for plane transitions. It was especially much lower than the STD algorithm, on which a discontinuity of the vortex is observed. However, the accuracy of the uni-directional interpolation schemes, commonly used in the literature, strongly alters the accuracy of the algorithm for arbitrary transitions and, in the end, the DC2 algorithm proves to be globally more accurate. \\

In view of the prohibitive cost of complex three-dimensional high-order interpolations, the choice of uni-directional interpolation schemes remains wise for cell-vertex algorithms. Meshes must be designed considering these results, especially when well-developed wakes cross refinement interfaces: with the DC algorithms, the use of plane transitions away from solids boundaries remains to be preferred.\\

Globally, it is shown that DC algorithms remain stable for high Reynolds number flows and greatly improve the accuracy of the grid interface. Moreover, the DC2 algorithm turns out to be the most accurate one for the realistic case of a turbulent cylinder wake flow. These last results may be due to reduced aliasing effects, which are all the more important as turbulent structures are under-resolved.\\

Furthermore, results obtained with the H-RR model on the vortex convection as well as on the cylinder wake flow, confirm the generic nature of our first study \cite{Astoul2020}, in which the STD algorithm was used. No harmful contribution of non-hydrodynamic modes at mesh transitions is noticeable with the DC algorithms. On the vortex case, neither spurious vorticity, nor any striations on the pressure field are indeed visible in the vicinity of the vortex. On the turbulent cylinder case, no high-frequency waves are observed.\\

A combination of the H-RR collision model (or any other model allowing for an efficient damping of non-hydrodynamic modes), and the DC2 algorithm seems to be a suitable choice for industrial aeroacoustic studies. As a perspective, it would then be interesting to evaluate the combination of these models on industrial aeronautical applications. 

\section*{Acknowledgements}
\label{acknowledgements}
The authors would like to gratefully acknowledge Gregoire Pont and Florian Renard for the fruitful discussions on linear system resolution. Acknowledgements are also expressed to Airbus Operations for HPC resources and ANRT/CIFRE for the financial support. This work has been carried out using ProLB, a Lattice-Boltzmann solver developed within a scientific collaboration including
CSSI, Renault, Airbus, Ecole Centrale de Lyon, CNRS and Aix-Marseille University

\newpage

\section*{Appendix A: Details on the cubic Mach correction term for the D3Q19 lattice}
\label{app:AppA}

This correction term $\psi$ allows to correct the deviation between the third-order moments obtained from the truncated equilibrium used for the D3Q19 given by the Eq.~(\ref{eq:fEq_herm}) and those obtained by calculating moments of the Boltzmann distribution function. These terms are given in Feng \textit{et al.} article \cite{Feng2019} and are recalled here:

\begin{equation} \label{eq:psi}
\begin{split}
	 \psi_i = \frac{\omega_i}{2c_s^4}  &\left[  \herm^{(2)}_{i, xx} \left(\frac{\partial}{\partial x} \Psi_{xxx} + \frac{\partial}{\partial y} \Psi_{xxy} + \frac{\partial}{\partial z} \Psi_{xxz}  \right) \right.\\ 
&	 \left.+ \herm^{(2)}_{i, yy} \left(\frac{\partial}{\partial x} \Psi_{xyy} + \frac{\partial}{\partial y} \Psi_{yyy} + \frac{\partial}{\partial z} \Psi_{yyz}  \right) \right.\\
&   \left.+ \herm^{(2)}_{i, zz} \left(\frac{\partial}{\partial x} \Psi_{xzz} + \frac{\partial}{\partial y} \Psi_{yzz} + \frac{\partial}{\partial z} \Psi_{zzz}  \right) \right.\\ 
&   \left.+ 2\herm^{(2)}_{i, xy} \left(\frac{\partial}{\partial x} \Psi_{xxy} + \frac{\partial}{\partial y} \Psi_{xyy} + \frac{\partial}{\partial z} \Psi_{xyz}  \right) \right.\\ 
&   \left.+ 2\herm^{(2)}_{i, xz} \left(\frac{\partial}{\partial x} \Psi_{xxz} + \frac{\partial}{\partial y} \Psi_{xyz} + \frac{\partial}{\partial z} \Psi_{xzz}  \right) \right.\\ 
&   \left.+ 2\herm^{(2)}_{i, yz} \left(\frac{\partial}{\partial x} \Psi_{xyz} + \frac{\partial}{\partial y} \Psi_{yyz} + \frac{\partial}{\partial z} \Psi_{yzz}  \right) \right],
\end{split}
\end{equation}

\noindent with $\Psi_{\alpha\beta\gamma}$ the deviation terms that can be written as follow in the isothermal approximation.

\begin{equation} \label{eq:Psi}
\begin{split}
&	 \Psi_{\alpha\alpha\alpha} = \rho u_\alpha ^3	,\\ 
&	 \Psi_{\alpha\alpha\beta} = \frac{1}{2}\rho u_\beta u_\gamma ^2,\\
&    \Psi_{\alpha\beta\gamma} = \rho u_\alpha u_\beta u_\gamma,
\end{split}
\end{equation}

\noindent These terms are estimated with second-order centered finite difference scheme as

\begin{align}
     \frac{\partial}{\partial i} \Psi_{\alpha\beta\gamma} \simeq  \frac{\Psi_{\alpha\beta\gamma} (\boldsymbol{x} + \boldsymbol{e_i}) - \Psi_{\alpha\beta\gamma} (\boldsymbol{x} - \boldsymbol{e_i})}{2},
    \label{eq:PSI_FD}
\end{align}

where $\boldsymbol{e_i} \in \{ \boldsymbol{e_x}, \boldsymbol{e_y}, \boldsymbol{e_z} \}$ is a unitary vector of the Cartesian coordinate system.

\section*{Appendix B: Examples of two-dimensional refinement interface with corners}
\label{app:AppB}

\noindent In this appendix, the particular example of a plane refinement interface illustrated in Fig.~\ref{fig:scheme_DC} is extended to interfaces with concave and convex corners. With the formalism described in the paper, the only difference between a plane interface and corners lies in the value of the velocity indexes which are associated with the sets $\mathcal{P},\mathcal{Q}^f,\mathcal{M}^f$. These sets determine the allowable values of the parameters $\Gamma_i$ and $\gamma_i$ as described in Table~\ref{tab:choice_Gamma_fine}.\\

\noindent The configuration of planar, concave and convex interfaces are the only three ones that can locally be found with Cartesian meshes in two dimensions. An extension to the three-dimensional case is straightforward by further distinghishing edges and corners. Anyway, the computation of the Jacobian matrix involved in Eq.~(\ref{eq:system_fi}) is done only once in advance and considers any possible shape for the interface. The choice of sets $\mathcal{P}$, $\mathcal{Q}^f$, $\mathcal{M}^f$ is directly substituted in the LBM code.\\

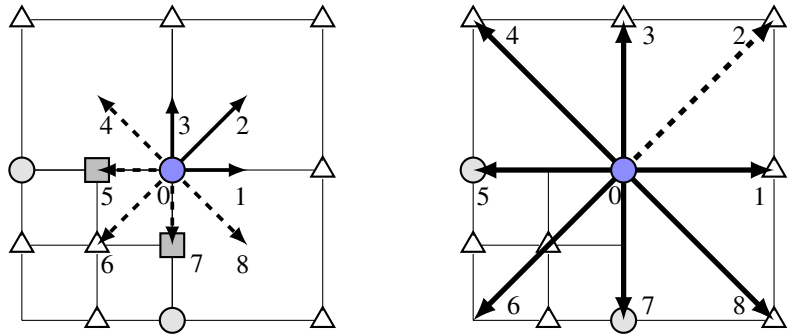
\begin{figure}[H]
		\begin{tikzpicture}[scale=1.]
			\draw (0,2) grid[step=2.] (2,4);
			\draw (2,0) grid[step=2.] (4,4);
			\draw (0,0) grid[step=1.] (2,2);
			\tikzstyle{TC}=[circle,draw,thick,fill=gray!22]
			\tikzstyle{TC0}=[circle,draw,thick,fill=blue!45]
			\tikzstyle{IE}=[rectangle,draw,thick,fill=gray!50,scale=1.35]
			\tikzstyle{IF}=[draw,rectangle,thick,fill=white]
			\tikzstyle{Mid}=[mark size=5pt,fill=white,thick]

			\draw (2,0) node[TC]{};
			\draw (0,2) node[TC]{};
			\draw (1,2) node[IE]{};
			\draw (2,1) node[IE]{};
			\node[Mid] at (1,0) {\pgfuseplotmark{triangle}};
			\node[Mid] at (1,1) {\pgfuseplotmark{triangle}};
			\node[Mid] at (0,1) {\pgfuseplotmark{triangle}};
			\node[Mid] at (4,0) {\pgfuseplotmark{triangle}};
			\node[Mid] at (4,2) {\pgfuseplotmark{triangle}};
			\node[Mid] at (4,4) {\pgfuseplotmark{triangle}};
			\node[Mid] at (0,4) {\pgfuseplotmark{triangle}};
			\node[Mid] at (2,4) {\pgfuseplotmark{triangle}};
			
			\draw [-> , >=latex ,line width=0.5mm] (2,2) -- (3,3);
			\draw [-> , >=latex ,line width=0.5mm] (2,2) -- (3,2);
			\draw [-> , >=latex ,line width=0.5mm,dashed] (2,2) -- (3,1);
			\draw [-> , >=latex ,line width=0.5mm,dashed] (2,2) -- (2,1);
			\draw [-> , >=latex ,line width=0.5mm] (2,2) -- (2,3);
			\draw [-> , >=latex ,line width=0.5mm,dashed] (2,2) -- (1,3);
			\draw [-> , >=latex ,line width=0.5mm,dashed] (2,2) -- (1,2);
			\draw [-> , >=latex ,line width=0.5mm,dashed] (2,2) -- (1,1);	
			\draw (2,2) node[TC0]{};		

 			\node[text width=6.5cm] at(5.05,1.67){0};
 			\node[text width=6.5cm] at(5.33,2.6){3};
 			\node[text width=6.5cm] at(5.5,0.75){7};
 			\node[text width=6.5cm] at(6.1,1.65){1};
 			\node[text width=6.5cm] at(6.1,2.6){2};
 			\node[text width=6.5cm] at(6.1,0.75){8};
 			\node[text width=6.5cm] at(4.3,0.75){6};
  			\node[text width=6.5cm] at(4.3,1.65){5};
 			\node[text width=6.5cm] at(4.3,2.6){4};

			\draw (6,2) grid[step=2.] (8,4);
			\draw (8,0) grid[step=2.] (10,4);
			\draw (6,0) grid[step=1.] (8,2);
			
			\draw (8,0) node[TC]{};
			\draw (6,2) node[TC]{};		
			\node[Mid] at (10,0) {\pgfuseplotmark{triangle}};
			\node[Mid] at (10,2) {\pgfuseplotmark{triangle}};
			\node[Mid] at (10,4) {\pgfuseplotmark{triangle}};
			\node[Mid] at (7,0) {\pgfuseplotmark{triangle}};
			\node[Mid] at (7,1) {\pgfuseplotmark{triangle}};
			\node[Mid] at (6,1) {\pgfuseplotmark{triangle}};		
			\node[Mid] at (6,4) {\pgfuseplotmark{triangle}};		
			\node[Mid] at (8,4) {\pgfuseplotmark{triangle}};		
			
			\draw [-> , >=latex ,line width=0.7mm] (8,2) -- (6,4);		
			\draw [-> , >=latex ,line width=0.7mm] (8,2) -- (6,2);
			\draw [-> , >=latex ,line width=0.7mm] (8,2) -- (6,0);
			\draw [-> , >=latex ,line width=0.7mm] (8,2) -- (8,0);
			\draw [-> , >=latex ,line width=0.7mm] (8,2) -- (8,4);
			\draw [-> , >=latex ,line width=0.7mm] (8,2) -- (10,0);
			\draw [-> , >=latex ,line width=0.7mm] (8,2) -- (10,2);
			\draw [-> , >=latex ,line width=0.7mm,dashed] (8,2) -- (10,4);		
			\draw (8,2) node[TC0]{}; 
			
 			\node[text width=.5cm] at(8.05,1.67){0};
 			\node[text width=.5cm] at(8.5,3.8){3};
 			\node[text width=.5cm] at(8.5,0.2){7};
 			\node[text width=.5cm] at(10.,1.65){1};
 			\node[text width=.5cm] at(9.7,3.8){2};
 			\node[text width=.5cm] at(9.7,0.2){8};
 			\node[text width=.5cm] at(6.7,0.2){6};
  			\node[text width=.5cm] at(6.3,1.65){5};
 			\node[text width=.5cm] at(6.7,3.8){4};
			
		\end{tikzpicture}
	\caption{\label{fig:corner1} Two dimensional representation of a concave corner refinement interface. (\protect\ArrowDash): Unknown distribution functions after a streaming step, (\protect\Arrow): known distribution functions. Left: fine domain, right: coarse domain.}
\end{figure}

A concave corner is displayed on Fig.~\ref{fig:corner1}. In this situation, the sets $\mathcal{P}$, $\mathcal{Q}^f$, $\mathcal{M}^f$ take the following values: $\mathcal{P}=\{0,1,3\}$, $\mathcal{Q}^f=\{2\}$ and $\mathcal{M}^f=\{4,5,6,7,8\}$.

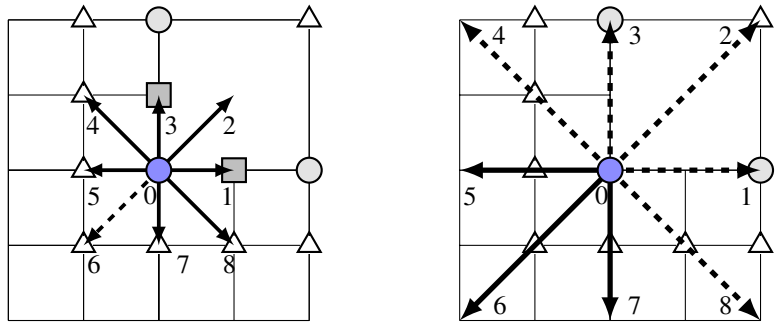
\begin{figure}[H]
		\begin{tikzpicture}[scale=1.]
			\draw (0,2) grid[step=2.] (2,4);
			\draw (2,0) grid[step=2.] (4,4);
			\draw (0,0) grid[step=1.] (2,2);
			\tikzstyle{TC}=[circle,draw,thick,fill=gray!22]
			\tikzstyle{TC0}=[circle,draw,thick,fill=blue!45]
			\tikzstyle{IE}=[rectangle,draw,thick,fill=gray!50,scale=1.35]
			\tikzstyle{IF}=[draw,rectangle,thick,fill=white]
			\tikzstyle{Mid}=[mark size=5pt,fill=white,thick]

 			\draw (2,2) grid[step=2.] (4,4);
			\draw (0,0) grid[step=1.] (2,4);
			\draw (2,0) grid[step=1.] (4,2);
			
			\draw (4,2) node[TC]{};
			\draw (2,4) node[TC]{};
			\draw (3,2) node[IE]{};
			\draw (2,3) node[IE]{};
 			\node[Mid] at (4,4) {\pgfuseplotmark{triangle}};
			\node[Mid] at (1,1) {\pgfuseplotmark{triangle}};
			\node[Mid] at (1,2) {\pgfuseplotmark{triangle}};
			\node[Mid] at (1,3) {\pgfuseplotmark{triangle}};
			\node[Mid] at (1,4) {\pgfuseplotmark{triangle}};
			\node[Mid] at (2,1) {\pgfuseplotmark{triangle}};
			\node[Mid] at (3,1) {\pgfuseplotmark{triangle}};
			\node[Mid] at (4,1) {\pgfuseplotmark{triangle}};
			
			\draw [-> , >=latex ,line width=0.5mm] (2,2) -- (3,3);
			\draw [-> , >=latex ,line width=0.5mm] (2,2) -- (3,2);
			\draw [-> , >=latex ,line width=0.5mm] (2,2) -- (3,1);
			\draw [-> , >=latex ,line width=0.5mm] (2,2) -- (2,1);
			\draw [-> , >=latex ,line width=0.5mm] (2,2) -- (2,3);
			\draw [-> , >=latex ,line width=0.5mm] (2,2) -- (1,3);
			\draw [-> , >=latex ,line width=0.5mm] (2,2) -- (1,2);
			\draw [-> , >=latex ,line width=0.5mm,dashed] (2,2) -- (1,1);	
			\draw (2,2) node[TC0]{};	

 			\node[text width=6.5cm] at(5.05,1.67){0};
 			\node[text width=6.5cm] at(5.33,2.6){3};
 			\node[text width=6.5cm] at(5.5,0.75){7};
 			\node[text width=6.5cm] at(6.1,1.65){1};
 			\node[text width=6.5cm] at(6.1,2.6){2};
 			\node[text width=6.5cm] at(6.1,0.75){8};
 			\node[text width=6.5cm] at(4.3,0.75){6};
  			\node[text width=6.5cm] at(4.3,1.65){5};
 			\node[text width=6.5cm] at(4.3,2.6){4};

 			\draw (8,2) grid[step=2.] (10,4);
			\draw (6,0) grid[step=1.] (8,4);
			\draw (8,0) grid[step=1.] (10,2);
			
			\draw (10,2) node[TC]{};
			\draw (8,4) node[TC]{};
 			\node[Mid] at (10,4) {\pgfuseplotmark{triangle}};
			\node[Mid] at (7,1) {\pgfuseplotmark{triangle}};
			\node[Mid] at (7,2) {\pgfuseplotmark{triangle}};
			\node[Mid] at (7,3) {\pgfuseplotmark{triangle}};
			\node[Mid] at (7,4) {\pgfuseplotmark{triangle}};
			\node[Mid] at (8,1) {\pgfuseplotmark{triangle}};
			\node[Mid] at (9,1) {\pgfuseplotmark{triangle}};
			\node[Mid] at (10,1) {\pgfuseplotmark{triangle}};
			
			\draw [-> , >=latex ,line width=0.7mm,dashed] (8,2) -- (10,4);
			\draw [-> , >=latex ,line width=0.7mm,dashed] (8,2) -- (10,2);
			\draw [-> , >=latex ,line width=0.7mm,dashed] (8,2) -- (10,0);
			\draw [-> , >=latex ,line width=0.7mm] (8,2) -- (8,0);
			\draw [-> , >=latex ,line width=0.7mm,dashed] (8,2) -- (8,4);
			\draw [-> , >=latex ,line width=0.7mm,dashed] (8,2) -- (6,4);
			\draw [-> , >=latex ,line width=0.7mm] (8,2) -- (6,2);
			\draw [-> , >=latex ,line width=0.7mm] (8,2) -- (6,0);	
			\draw (8,2) node[TC0]{};	
			
 			\node[text width=.5cm] at(8.05,1.67){0};
 			\node[text width=.5cm] at(8.5,3.8){3};
 			\node[text width=.5cm] at(8.5,0.2){7};
 			\node[text width=.5cm] at(10.,1.65){1};
 			\node[text width=.5cm] at(9.7,3.8){2};
 			\node[text width=.5cm] at(9.7,0.2){8};
 			\node[text width=.5cm] at(6.7,0.2){6};
  			\node[text width=.5cm] at(6.3,1.65){5};
 			\node[text width=.5cm] at(6.7,3.8){4};
		\end{tikzpicture}
	\caption{\label{fig:corner2} Two dimensional representation of a convex corner refinement interface. (\protect\ArrowDash): Unknown distribution functions after a streaming step, (\protect\Arrow): known distribution functions. Left: fine domain, right: coarse domain.}
\end{figure}

A convex corner is displayed on Fig.~\ref{fig:corner2}. This time, the sets $\mathcal{P}$, $\mathcal{Q}^f$, $\mathcal{M}^f$ take the following values: $\mathcal{P}=\{0,5,7\}$ , $\mathcal{Q}^f=\{1,2,3,4,8\}$ and $\mathcal{M}^f=\{6\}$.

Coarse distributions function are all reconstructed with Eq.~(\ref{eq:rescfc}) whatever the shape of the interface.

\newpage

\bibliographystyle{unsrt}
\bibliography{refs}

\begin{thebibliography}{10}
\bibitem{Lapack}
Lapack is a software package provided by univ. of tennessee; univ. of
  california, berkeley; univ. of colorado denver; and nag ltd.
\end{thebibliography}

\end{document}